\documentclass[pre,aps,twocolumn,superscriptaddress,showpacs]{revtex4-1}
\usepackage{amssymb}
\usepackage{epsfig}
\usepackage{amsmath}
\usepackage{subfigure}
\usepackage{graphicx}
\usepackage{textcomp}
\usepackage{url}
\usepackage{float}
\usepackage{color}
\usepackage{cases}
\usepackage[normalem]{ulem}

\usepackage[colorlinks=true, urlcolor=blue, anchorcolor=blue, citecolor=blue,filecolor=blue,linkcolor=blue,menucolor=blue
]{hyperref}

\usepackage{color}

\usepackage{epsfig}
\usepackage{textcomp}
\usepackage{epstopdf}

\begin{document}

\title{Large fluctuations of a Kardar-Parisi-Zhang interface on a half-line}

\author{Baruch Meerson}
\email{meerson@mail.huji.ac.il}
\author{Arkady Vilenkin}
\email{vilenkin@mail.huji.ac.il}
\affiliation{Racah Institute of Physics, Hebrew University of
Jerusalem, Jerusalem 91904, Israel}


\begin{abstract}

Consider a stochastic interface $h(x,t)$, described by the $1+1$ Kardar-Parisi-Zhang (KPZ) equation on the half-line $x\geq 0$. The interface is initially flat, $h(x,t=0)=0$, and driven by a Neumann boundary condition $\partial_x h(x=0,t)=A$ and by the noise. We study the short-time probability distribution $\mathcal{P}\left(H,A,t\right)$ of the one-point height $H=h(x=0,t)$. Using the optimal fluctuation method, we show that $-\ln \mathcal{P}\left(H,A,t\right)$ scales as $t^{-1/2} s \left(H,A t^{1/2}\right)$. For small and moderate $|A|$
this more general scaling reduces to the familiar simple scaling $-\ln \mathcal{P}\left(H,A,t\right)\simeq t^{-1/2} s(H)$, where $s$ is independent of $A$ and time and equal to one half of the corresponding large-deviation function for the full-line problem. For large $|A|$ we uncover two asymptotic regimes. At very short time the simple scaling is restored, whereas at intermediate times the scaling remains more general and $A$-dependent. The distribution tails, however, always exhibit the simple scaling in the leading order.

\end{abstract}

\maketitle

\section{Introduction}

The Kardar-Parisi-Zhang (KPZ) equation \citep{KPZ} describes non-equilibrium stochastic dynamics of the height $h(x,t)$ of a growing interface at the point $x$ of a substrate at time $t$:
\begin{equation}
\label{eq:KPZ_dimensional}
\partial_{t}h=\nu\partial_{x}^{2}h+\frac{\lambda}{2}\left(\partial_{x}h\right)^{2}+\sqrt{D}\,\xi(x,t).
\end{equation}
Here $\xi(x,t)$ is a Gaussian noise with zero average and
\begin{equation}\label{correlator}
\langle\xi(x_{1},t_{1})\xi(x_{2},t_{2})\rangle = \delta(x_{1}-x_{2})\delta(t_{1}-t_{2}).
\end{equation}
Throughout this paper we assume, without loss of generality,  that the symmetry-breaking nonlinearity coefficient $\lambda<0$  \citep{signlambda}.
The KPZ dynamics in 1+1 dimension have been studied in great detail. At late times, the characteristic width  of the KPZ interface increases as $t^{1/3}$, and the lateral correlation length increases as $t^{2/3}$.
The exponents $1/3$ and $2/3$ define an important universality class of the 1+1 dimensional non-equilibrium growth \cite{HHZ,Barabasi,Krug,Corwin,QS,S2016,Takeuchi2017}.   Among the more detailed characterizations of the KPZ growth is the full probability distribution $\mathcal{P}\left(H,t\right)$ of the interface height at a given space-time point: $H=h\left(x=0,t\right)$. Remarkably, the form of this distribution, even at arbitrarily long times, depends on the initial shape of the interface $h\left(x,t=0\right)$, see Refs.  \cite{QS,S2016,Takeuchi2017} for recent reviews.

Traditionally, the interest in the KPZ equation and related models has been focused on their long-time dynamics and universality. With an emergence of interest in large deviations in stochastic systems out of equilibrium, there have been a growing number of recent studies
of short-time, $t \ll \nu^{5}/(D^{2}\lambda^{4})$, dynamics of the one-point height distribution $\mathcal{P}\left(H,t\right)$. This interest was sparked by a discovery of a new scaling behavior of the distribution, including its tails which describe large deviations of the height.  As of present, short-time height distributions are known exactly for the droplet \citep{DMRS}, stationary \citep{LeDoussal2017}  and flat \cite{SM2018} initial conditions. For several other initial conditions the leading-order asymptotics of the distribution tails  have been determined. Quite often the distribution tails, predicted at short times, hold (at sufficiently large $H$)  at arbitrary long times \cite{MKV}. For the droplet initial condition this important property is well established by now \cite{SMP,Corwinetal2018}.

Almost all of the previous works on the one-point height statistics assumed an infinite substrate:  $-\infty<x<\infty$. In such cases, $\mathcal{P}\left(H,t\right)$  at short times turns out to behave, in a proper moving frame \cite{footnote:displacement}, as $-\ln\mathcal{P}\simeq s \left(H\right)/\sqrt{t}$. Here the function $s(H)$ plays the role of the large  deviation function of the height fluctuations. Recently, the role of boundaries in the dynamics of  $\mathcal{P}\left(H,t\right)$ has attracted attention. Ref. \cite{SMS2018} studied the short-time
probability distribution of the KPZ height at one point on a ring of length $2L$, and the authors identified a ``phase diagram" of different scaling behaviors  of $\ln \mathcal{P}\left(H,L,t\right)$  in the $(L/\sqrt{t},H)$ plane. A more basic setting is the half-line $x \ge 0$, and there have been several studies dealing with it, both at long \cite{GueudreLeDoussal2012,Borodin2016,Barraquand2017,CorwinShen2018,ItoTakeuchi2018,Corwinetal2018,Krajenbrink2018} and at short \cite{GueudreLeDoussal2012,SM2018,Krajenbrink2018} times.

Here we will focus on the short-time regime. Smith and Meerson \cite{SM2018} employed the optimal fluctuation method (OFM, which we will briefly review toward the end of the Introduction) and established a simple relation between \emph{any} full-line problem with spatial mirror symmetry $x \leftrightarrow -x$ of the optimal path  and the corresponding half-line ($x \ge 0$) problem with the same initial condition and the homogeneous Neumann boundary condition $\partial_x h(0,t)=0$. The relation is
\begin{equation}\label{relation}
s \left(H\right)=\frac{1}{2} s_{\text{full}}\left(H\right) ,
\end{equation}
where $s\left(H\right)$  is the large deviation function for the half-line problem, and $s_{\text{full}}\left(H\right)$ is the large deviation function for the full-line problem.
Although Eq.~(\ref{relation}) is a simple consequence of the OFM formalism \cite{SM2018}, it is far from intuitive. Indeed, it implies that it is much more likely to observe unusually large values of $H$ in a half-line system than in the full-line system with otherwise the same parameters \cite{naive}.

Now let us consider a half-line problem with a more general Neumann boundary condition at $x=0$:
\begin{equation}\label{A}
\partial_x h(0,t) = A.
\end{equation}
For $A\neq 0$ this boundary condition drives the KPZ interface even in the absence of noise.
Recently Krajenbrink and Le Doussal \cite{Krajenbrink2018} (see also Ref. \cite{GueudreLeDoussal2012}) considered this problem for the droplet initial condition. They extracted the $t\to 0$ asymptotics from exact representations for $\mathcal{P}\left(H,A,t\right)$ for three particular values of $A$. These included $A=-\infty$, corresponding to the ``hard wall", $A=0$, corresponding to the ``reflecting wall", and a finite positive value of $A$, corresponding to the  so called critical case \cite{Kardar}. For all three values of $A$,  Krajenbrink and Le Doussal arrived at the scaling behavior $-\ln\mathcal{P}\simeq S \left(H\right)/\sqrt{t}$. For the latter two values of $A$, they observed that the $H\to \infty$ and $H\to -\infty$ tails of $\ln \mathcal{P}\left(H,A,t\right)$ in the reflecting and critical cases  coincide and obey Eq.~(\ref{relation}) \cite{Krajenbrink2018}. These findings suggest that, in the limit of $t\to 0$, a non-zero but finite $A$ does not affect the height statistics.  In the present work we show that this conjecture is correct for the flat initial condition. We also show that, for sufficiently large $|A|$, there is an additional asymptotic regime -- of intermediate times -- where $A$ is relevant, and where the scaling of $-\ln\mathcal{P}$ with time is different. The distribution tails, however, always exhibit the simple $A$-indendent scaling, up to subleading terms which violate it.

Our approach is based on the OFM (also known as the weak-noise theory, instanton method, and macroscopic fluctuation theory). The OFM originated in condensed matter physics \cite{Halperin,Langer,Lifshitz,Lifshitz1988} and found many applications in theory of turbulence and turbulent transport
\citep{turb1,turb2,turb3}, diffusive lattice gases \citep{bertini2015} and
stochastic reactions on lattices \citep{EK, MS2011}.
It has been employed in many studies of the KPZ equation and related systems \citep{Mikhailov1991, GurarieMigdal1996,Fogedby1998, Fogedby1999,Nakao2003, KK2007,KK2008,KK2009,Fogedby2009,MKV,KMSparabola,Janas2016, MeersonSchmidt2017, MSV_3d, SMS2018, SKM2018,SM2018}. The starting point of the OFM is the path integral of the stochastic process, conditioned on a specified large deviation. If the noise is effectively weak, the path integral can be evaluated using Laplace's method. This leads to a
variational problem, the solution of which is the most probable, or optimal, path of the stochastic process, and the most probable realization of the noise, conditioned on the specified large deviation. The variational problem can be formulated as a classical Hamiltonian field theory. The action, evaluated on the optimal path, yields $\mathcal{P}$ up to a pre-exponential factor. For a nonzero $A$, the OFM formalism yields a more general scaling behavior,
\begin{equation}\label{newscaling}
-\ln\mathcal{P}(H,A,t)\simeq \frac{s \left(H,A\sqrt{t}\right)}{\sqrt{t}} .
\end{equation}
and this work addresses the consequences of this fact.

The remainder  of the paper is organized as follows. In Sec.~\ref{expectedevolution} we consider the evolution of an initially flat KPZ interface in the absence of the noise. This evolution determines the \emph{expected} value of height $H$. In Sec.~\ref{sec:OFM} we present the OFM formalism and expose our analytical and numerical calculations of the function $s \left(H,A\sqrt{t}\right)$ in different regimes. 
Our results are summarized and briefly discussed in Sec. \ref{disc}.

\section{Deterministic evolution}
\label{expectedevolution}
\begin{figure}[ht]
\includegraphics[width=0.3\textwidth,clip=]{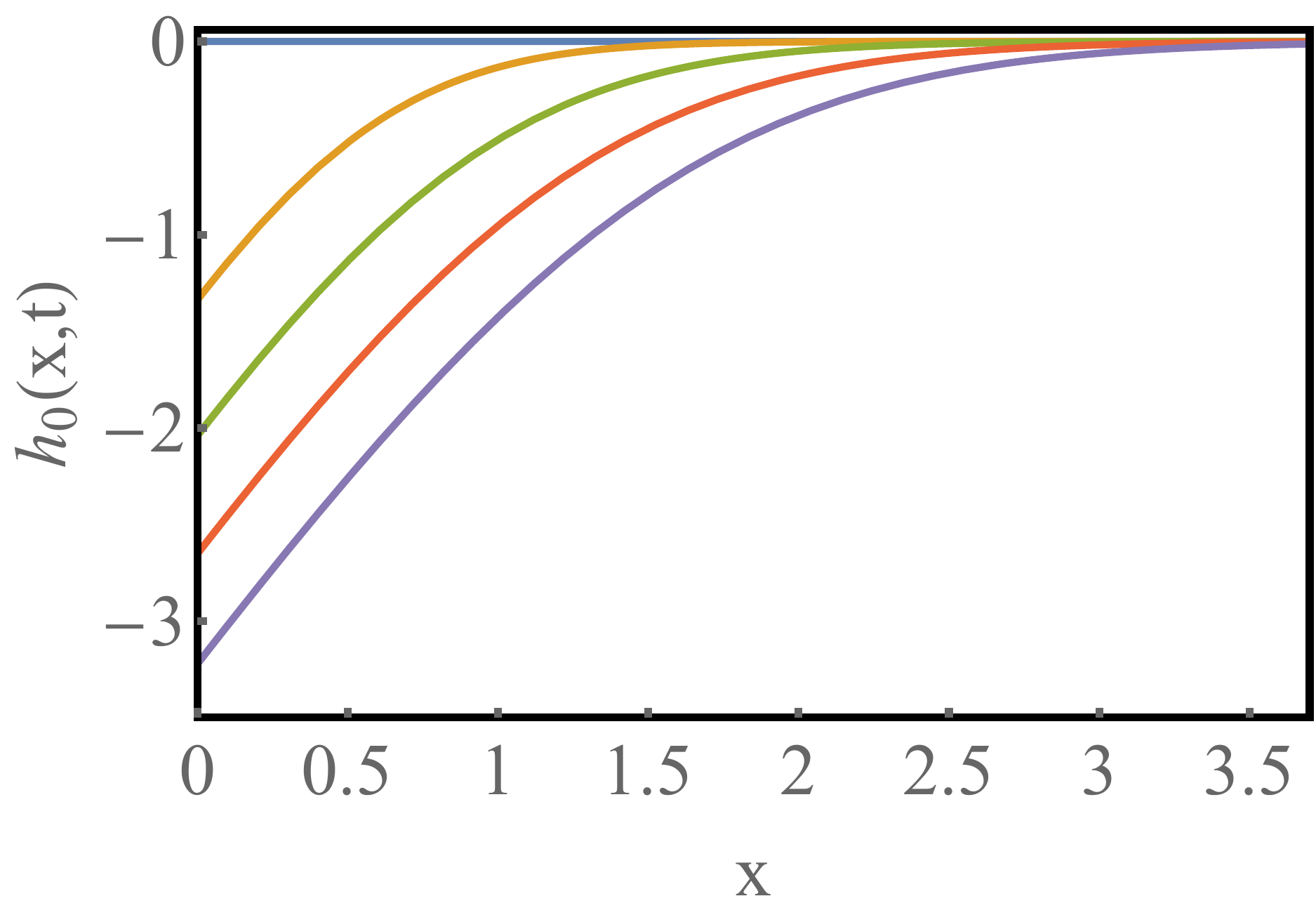}
\includegraphics[width=0.3\textwidth,clip=]{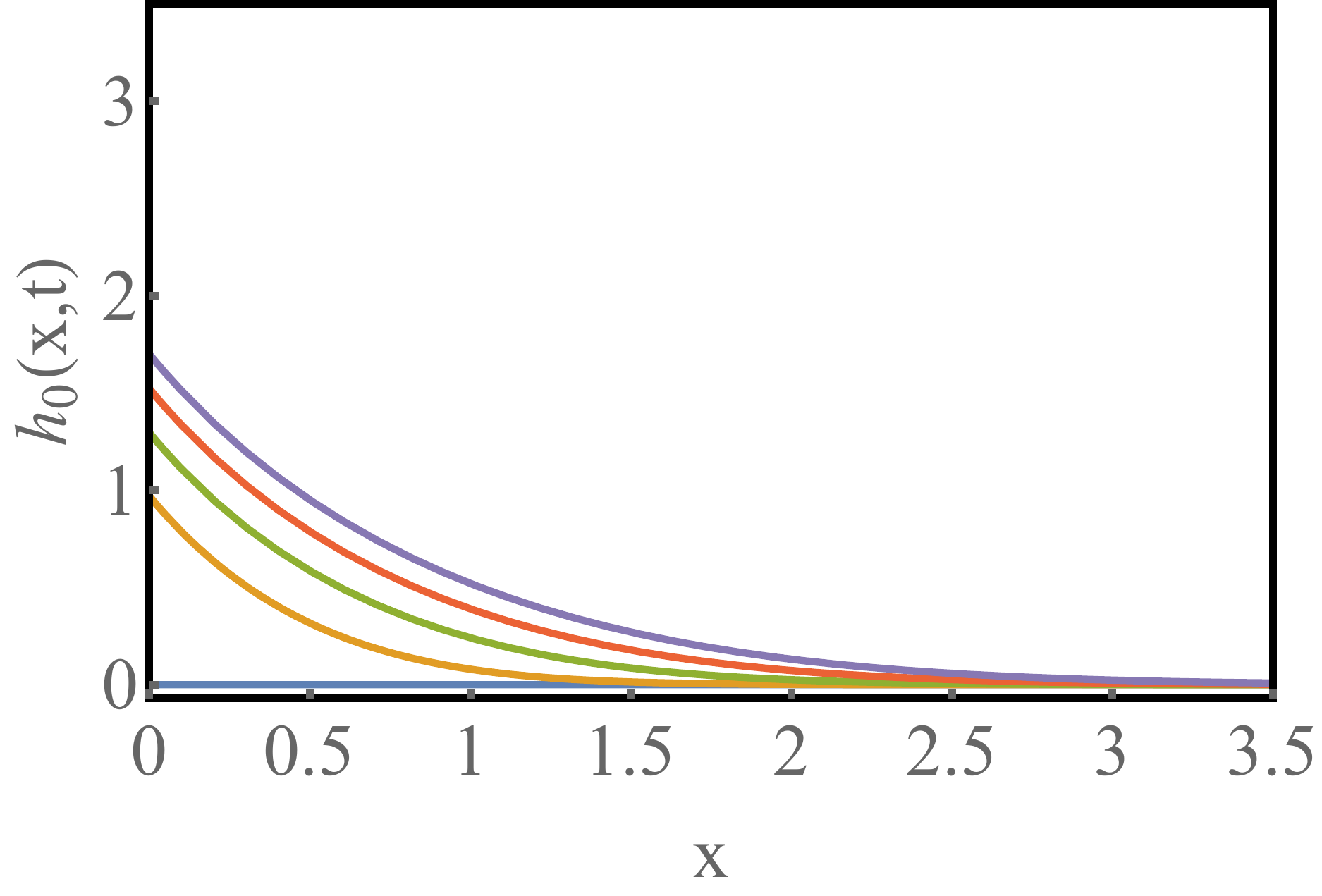}
\caption{Deterministic evolution of the interface as described by Eq.~(\ref{hdet})  for $a=2$ (upper panel) and $a=-2$ (lower panel) at times $0$, $1/4$, $1/2$, $3/4$ and $1$: from top to bottom in the upper panel, and from bottom to top in the lower one. The identical scales of the two panels emphasize the asymmetry of the deterministic solutions at positive and negative $a$.}
\label{detfig1}
\end{figure}

In the absence of noise, an initially flat interface will evolve if and only if $A\neq 0$. This evolution is described by the deterministic KPZ equation
\begin{equation}
\label{KPZdet}
\partial_{t}h=\nu\partial_{x}^{2}h+\frac{\lambda}{2}\left(\partial_{x}h\right)^{2}.
\end{equation}
Let the observation time be $T$. Upon rescaling  $t/T \to t$, $x/\sqrt{\nu T}\to x$ and $\left|\lambda\right|h/\nu\to h$  Eq.~(\ref{KPZdet}) becomes dimensionless,
\begin{equation}
\label{KPZdet1}
\partial_{t}h=\partial_{x}^{2}h-\frac{1}{2}\left(\partial_{x}h\right)^{2}.
\end{equation}
Importantly, the interface slope at $x=0$ undergoes a $T$-dependent rescaling:
\begin{equation}\label{AA}
\partial_x h(0,t) = a,     \quad\mbox{where}\quad a = \frac{|\lambda|\sqrt{T}}{\sqrt{\nu}} \,A.
\end{equation}
Equation~(\ref{KPZdet1}) should be solved with the boundary condition  (\ref{AA}) and the initial condition $h(x,0)=0$. The Cole-Hopf ansatz $h(x,t)=-2 \ln u(x,t)$
transforms Eq.~(\ref{KPZdet1}) into the diffusion equation $\partial_t u =\partial_{x}^2 u$ \cite{Whitham}. The Neumann condition (\ref{AA}) becomes a Robin condition
\begin{equation}\label{robin}
\partial_x u(0,t) + \frac{a}{2}\,u(0,t) = 0.
\end{equation}
The solution, in terms of $h(x,t)$, is
\begin{equation}\label{hdet}
  h_0(x,t) = -2 \ln \left[e^{\frac{a}{4}(a t-2 x)}
   \text{erfc}\left(\frac{x-a t}{2
   \sqrt{t}}\right)+\text{erf}\left(\frac{x}{2
   \sqrt{t}}\right)\right] ,
\end{equation}
where $\text{erf}\,z = (2/\sqrt{\pi})\int_0^z e^{-\xi^2}\,d\xi$ is the error function, and $\text{erfc}\,z =1-\text{erf} \,z$. Figure \ref{detfig1} shows the deterministic (that is, expected) time history of the interface for $a=2$ and $a=-2$.

\begin{figure}[ht]
\includegraphics[width=0.3\textwidth,clip=]{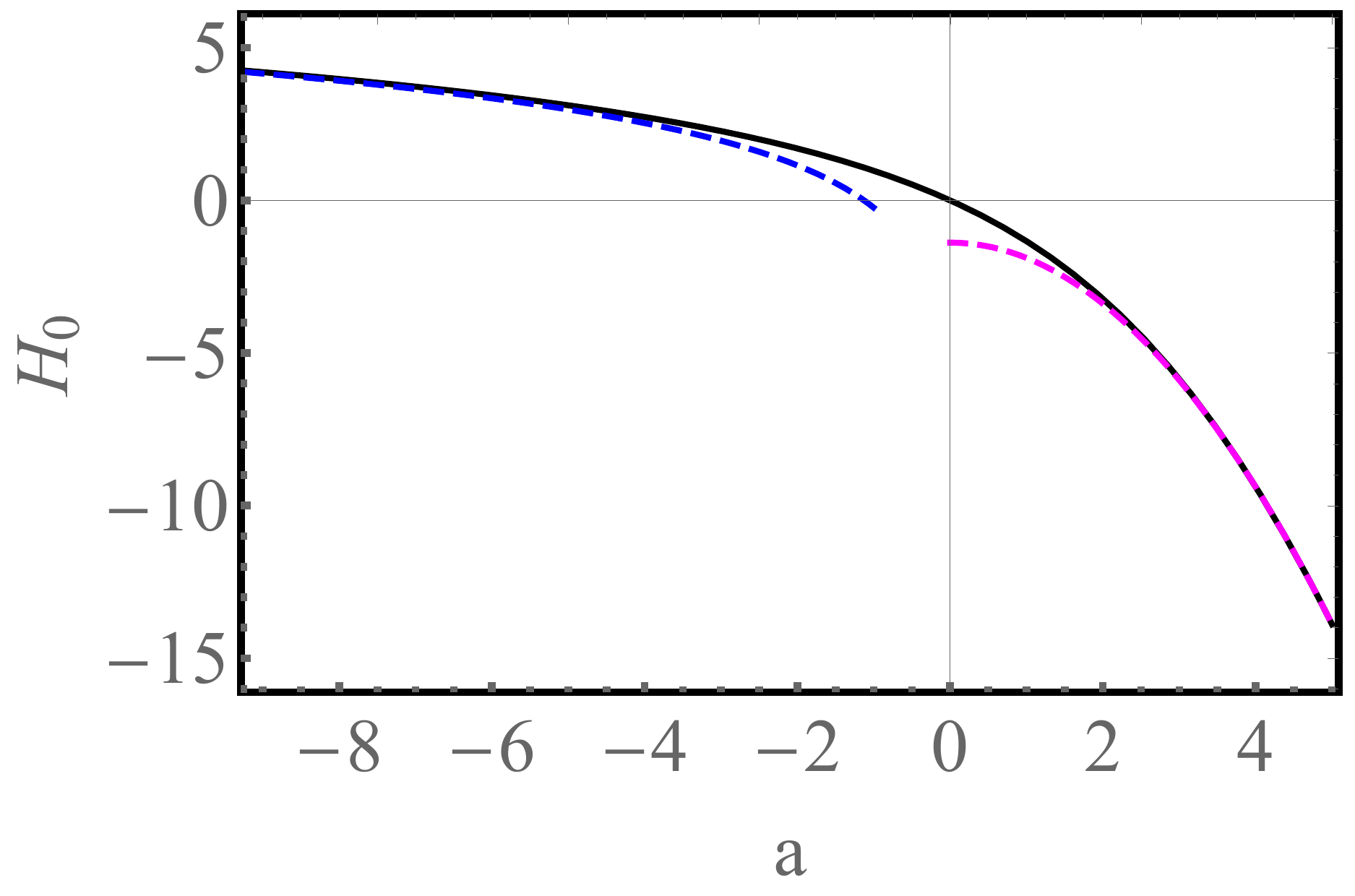}
\caption{The expected one-point interface height $H_0=h(0,1)$ versus $a$. Solid line: Eq.~(\ref{H0}), dashed lines: asymptotics (\ref{H0asymp}).}
\label{H0fig}
\end{figure}

The expected interface height at $x=0$ and $t=1$ is
\begin{equation}\label{H0}
h_0(0,1)\equiv H_0(a) = -2 \ln \left[e^{\frac{a^2}{4}}
   \text{erfc}\left(-\frac{a}{2}\right)\right] .
\end{equation}
The function of $H_0(a)$ vanishes at $a=0$ and is strongly asymmetric with respect to $a=0$, see Fig.~\ref{H0fig}. The  asymptotes of $H_0(a)$ are the following:
\begin{equation}
\label{H0asymp}
H_0 (a)=\begin{cases}
-\frac{a^2}{2}-2 \ln 2 +\dots, & a\gg 1 ,\\
\ln \left(\frac{\pi  a^2}{4}\right)+\dots, & -a \gg 1 ,
\end{cases}
\end{equation}

\begin{figure}[ht]
\includegraphics[width=0.3\textwidth,clip=]{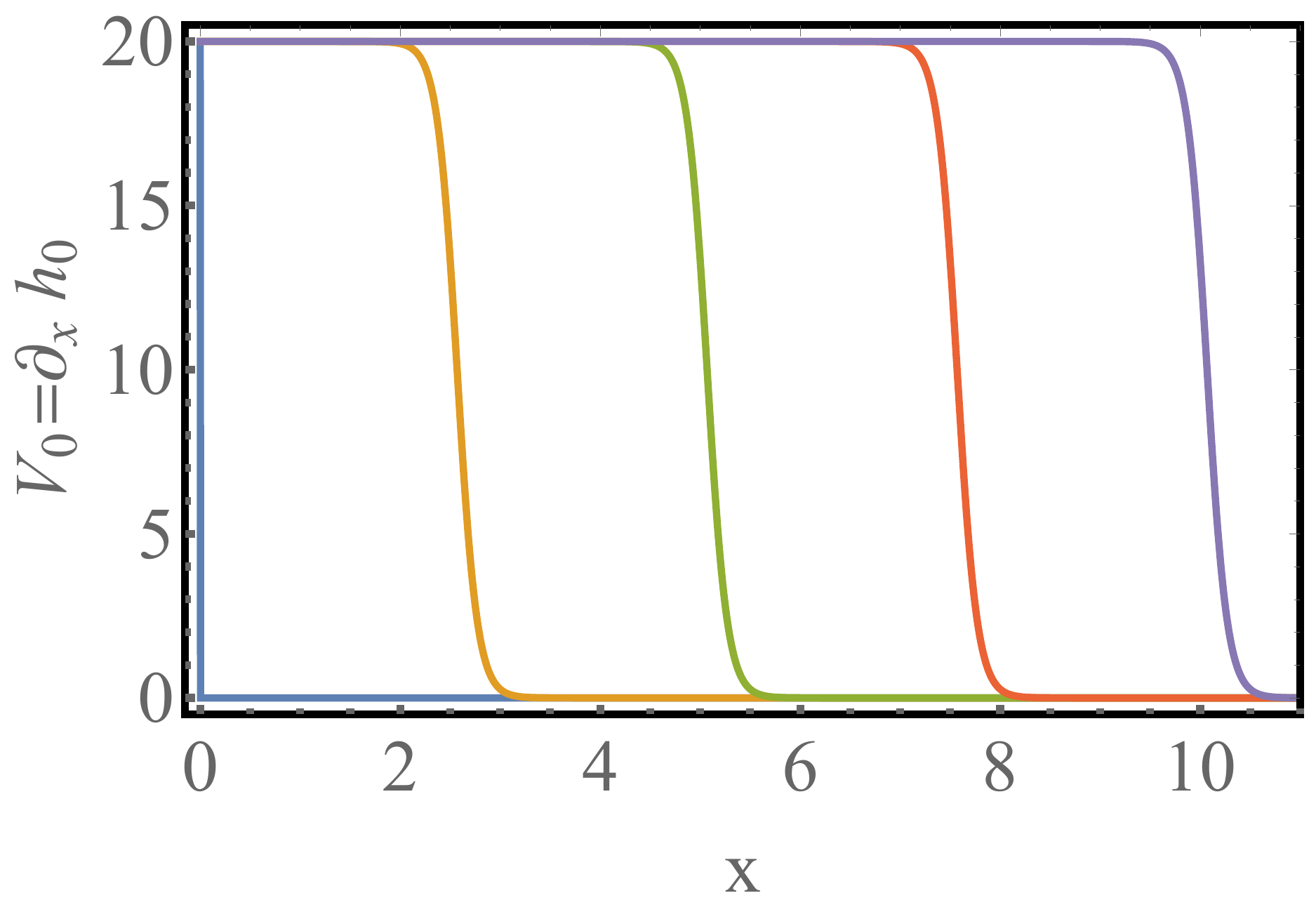}
\caption{The expected interface slope $V_0(x,t)=\partial_x h_0(x,t)$ versus $x$ for $a=20$ at $t=0$, $1/4$, $1/2$, $3/4$ and $1$ (from left to right), as described by Eq.~(\ref{Vdet}).}
\label{shock}
\end{figure}

We will also need to know the evolution of the expected interface slope $V_0(x,t)=\partial_x h_0(x,t)$:
\begin{equation}\label{Vdet}
V_0(x,t)=\frac{a \,e^{\frac{a}{4} (a t-2 x)}
   \text{erfc}\left(\frac{x-a t}{2
   \sqrt{t}}\right)}{e^{\frac{a}{4} (a t-2 x)}
   \text{erfc}\left(\frac{x-a t}{2
   \sqrt{t}}\right)+\text{erf}\left(\frac{x}{2
   \sqrt{t}}\right)}.
\end{equation}
The deterministic solution simplifies for $a\gg 1$. In this limit Eq.~(\ref{Vdet}) describes the formation and propagation of a simple Burgers shock \cite{Whitham} with velocity $a/2$, where $V_0$ ``jumps" from $V_0=a$ behind the shock to $V_0=0$ in front of the shock. The width of the transition region is of order of $1$, see Fig. \ref{shock}.

\section{Distribution variance and tails}
\label{sec:OFM}

\subsection{OFM formulation}
Now we return to the stochastic KPZ equation~(\ref{eq:KPZ_dimensional}). After the rescaling, described above, it becomes
\begin{equation}
\label{eq:KPZ_dimensionless}
\partial_{t}h=\partial_{x}^{2}h-\frac{1}{2}\left(\partial_{x}h\right)^{2}+\sqrt{\epsilon} \, \xi\left(x,t\right),
\end{equation}
where $\epsilon=D\lambda^{2}\sqrt{T}/\nu^{5/2}$ is the dimensionless noise magnitude. In the short-time limit, $\epsilon \to 0$, one can evaluate the path integral, corresponding to Eq.~(\ref{eq:KPZ_dimensionless}), by Laplace's method. This procedure boils down to
a minimization problem for the action
\begin{equation}
\label{eq:sdyn_def}
s =
\frac{1}{2}\int_{0}^{1}dt\int_{0}^{\infty}dx
\left[\partial_{t}h-\partial_{x}^{2}h+\frac{1}{2}\left(\partial_{x}h\right)^{2}\right]^{2}.
\end{equation}
The ensuing Euler-Lagrange equation can be recast into Hamiltonian equations for the optimal history of interface  $h\left(x,t\right)$ and its canonically conjugate ``momentum'' $\rho\left(x,t\right)$ which describes the optimal realization of the noise $\xi(x,t)$ \citep{Fogedby1998, KK2007, MKV}:
\begin{eqnarray}
  \partial_{t} h &=& \delta \mathcal{H}/\delta \rho = \partial_{x}^2 h - \frac{1}{2} \left(\partial_x h\right)^2+\rho ,  \label{eqh}\\
  \partial_{t}\rho &=& - \delta \mathcal{H}/\delta h = - \partial_{x}^2 \rho - \partial_x \left(\rho \partial_x h\right). \label{eqrho}
\end{eqnarray}
Here
\begin{equation*}
\mathcal{H}=\int_{0}^{\infty}\!dx\,\rho\left[\partial_{x}^{2}h- \frac{1}{2}\left(\partial_{x}h\right)^{2}+\rho/2\right]
\end{equation*}
is the Hamiltonian. One boundary condition at $x=0$ is the fixed slope condition~(\ref{AA}). The additional condition at $x=0$  is
\begin{equation}\label{BCrho}
\partial_x \rho(0,t) +a \rho(0,t) = 0.
\end{equation}
This zero-flux condition ensures  that the boundary term at $x=0$, coming from the integration by parts of the linear variation of the action, vanishes as it should. The initial condition is
\begin{equation}
\label{eq:flat_IC}
h \left(x,t=0\right)=0,\quad x\geq 0.
\end{equation}
The condition
\begin{equation}
\label{eq:h_1_H}
h(x=0,t=1)=H
\end{equation}
can be translated into a ``final" condition for $\rho$ \citep{KK2007}:
\begin{equation}
\label{pT}
\rho\left(x,t=1\right)=\Lambda\,\delta\left(x\right),
\end{equation}
with a Lagrange multiplier $\Lambda$, ultimately determined by $H$.

Once the OFM problem is solved, we can evaluate the action $s$:
\begin{equation}
\label{eq:sdyn_recast}
s=s(H,a)=\frac{1}{2}\int_{0}^{1}dt\int_{0}^{\infty}dx\,\rho^{2}\left(x,t\right).
\end{equation}
Up to a small correction, $-\ln \mathcal{P} \simeq s(H,a)/\epsilon$, or
\begin{equation}
-\ln\mathcal{P}\left(H,T,A\right)
\simeq\frac{\nu^{5/2}}{D\lambda^{2}\sqrt{T}}\,\,s\left(\frac{\left|\lambda\right|H}{\nu},
\frac{A|\lambda|\sqrt{T}}{\sqrt{\nu}}\right)
 \label{actiondgen}
\end{equation}
in the dimensional variables, as announced in Eq.~(\ref{newscaling}).  Now we see
that three different regimes are possible:
\begin{enumerate}
\item{For very short observation times,
\begin{equation}\label{veryshort}
T\ll \frac{\nu}{A^2\lambda^2} ,
\end{equation}
the second argument of the function $s$ in Eq.~(\ref{actiondgen}) can be sent to zero. In this limit $-\ln \mathcal{P}$ becomes independent of $A$ and exhibits the simple scaling
   $s(H)/\sqrt{t}$.}
\item{For intermediate observation times,
\begin{equation}\label{interm}
\frac{\nu}{A^2\lambda^2}\lesssim T \ll \frac{\nu^5}{D^2 \lambda^4},
\end{equation}
the OFM is still applicable for the whole height distribution, but a nontrivial scaling of $-\ln \mathcal{P}$ with time can appear. This regime is possible only
when $|A|$ is much larger than an intrinsic height gradient scale of the KPZ equation:
\begin{equation}\label{largeA}
\frac{|A|\nu^2}{D|\lambda|}\gg 1.
\end{equation}}
\item{For longer observation times,
\begin{equation}\label{long}
T \gtrsim\frac{\nu^5}{D^2 \lambda^4},
\end{equation}
the OFM is inapplicable for typical fluctuations of the height, but may still be applicable in the tails.}
\end{enumerate}
Here we only consider the regimes 1 and 2. We will see later that, in the distribution tails, $H\to \pm \infty$,  the dependence of the function $s$ on its  second argument appears only in a subleading order, and the simple scaling  $-\ln \mathcal{P}\simeq s(H)/\sqrt{t}$ is observed up to subleading corrections.

For $A=0$, when $H_0=0$, the short-time large-deviation function $s(H)$ for the flat initial condition is known exactly \cite{SM2018}, and it obeys the relation~(\ref{relation}).
In its turn, $s_{\text{full}} (H)$ has been recently found in Ref. \cite{SM2018} by (i) exploiting, in the OFM formalism, a non-trivial symmetry of the KPZ equation in $1+1$ dimension \cite{Frey1996,Canet2011,Mathey2017}, (ii) establishing a simple mapping between the OFM problems with flat and stationary initial conditions, and (iii) using exact short-time results, extracted in Ref. \cite{LeDoussal2017} from the known exact representation for the stationary case \cite{IS,Borodinetal}. For further reference, we present asymptotics of $s(H,A=0)$ for the half-line problem:
\begin{equation}
\label{forreference}
s\!=\!\begin{cases}
\frac{4\sqrt{2}}{15\pi}H^{5/2}+\frac{4\sqrt{2}}{3\pi}H^{3/2}\ln H\\
+\frac{2\sqrt{2}}{9\pi}\left[2+3\ln\left(\frac{4}{9\pi^{2}}\right)\right]H^{3/2}+\dots, & H\to+\infty,\\
\sqrt{\!\frac{\pi}{8}}\,H^{2}+\sqrt{\!\frac{\pi}{288}}\,\left(\pi-3\right)H^{3}+\dots, & \left|H\right|\ll1,\\
\frac{4\sqrt{2}}{3}\!\left|H\right|^{3/2}\!\!-\!4\sqrt{2}\,\ln\!\left(2\right)\!\left|H\right|^{1/2}\!+\dots , & H\to-\infty .
\end{cases}
\end{equation}
The same results are obtained when $A\neq 0$, but the observation time $T$ is very short, see above. This completes our consideration of the limit $T\to 0$. In the remainder of the paper we will focus on the regime of large $|A|$ and intermediate times, see Eqs.~(\ref{interm}) and~(\ref{largeA}), where $s(H,a)$ is unknown. In the absence of exact solution of the OFM problem our strategy will be similar to that of the previous works on short-time large deviations of KPZ interfaces \citep{KK2007,KK2008,KK2009,MKV,KMSparabola,Janas2016,
MeersonSchmidt2017, SKM2018}. We will employ three different perturbation approaches: to obtain the leading (and sometimes even subleading) asymptotics for the left and right tails of $\mathcal{P}(H)$, and also to evaluate the variance of $\mathcal{P}(H)$ for $a\gg 1$, when the left inequality sign in the double inequality~(\ref{interm}) becomes $\ll$.
Finally, we will solve the OFM problem numerically \cite{numerics},  find $s(H,a)$ in different parameter regimes and verify our approximate analytical results.

\subsection{Variance}

\begin{figure}[ht]
\includegraphics[width=0.3\textwidth,clip=]{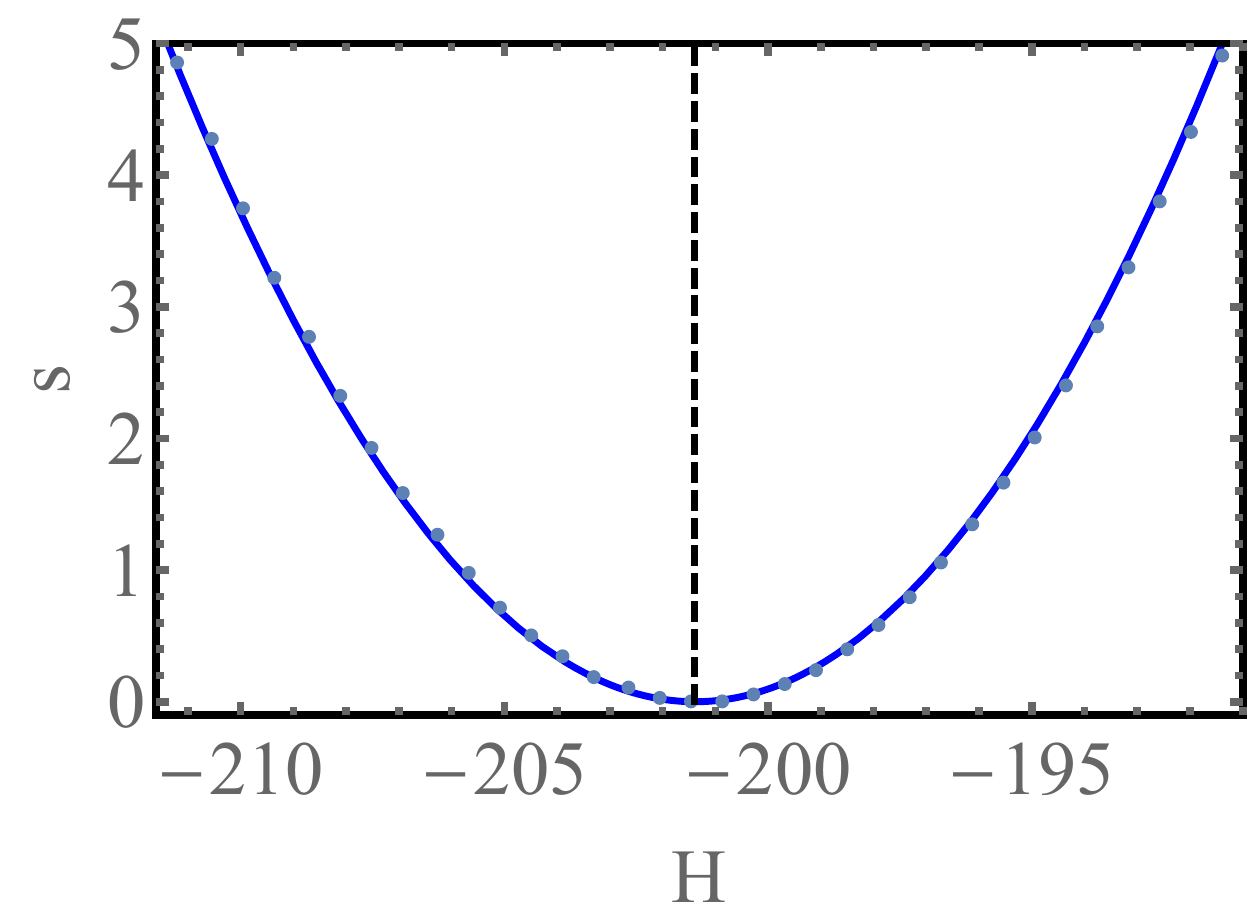}
\caption{The Gaussian action $s(H,a=20)$, predicted by Eq.~(\ref{varlargeA}) (solid line), is compared with the action computed numerically for $a=20$ (symbols). The dashed line shows the expected height $H_0(a=20)\simeq -201.4$.}
\label{20linear}
\end{figure}

\begin{figure}[ht]
\includegraphics[width=0.3\textwidth,clip=]{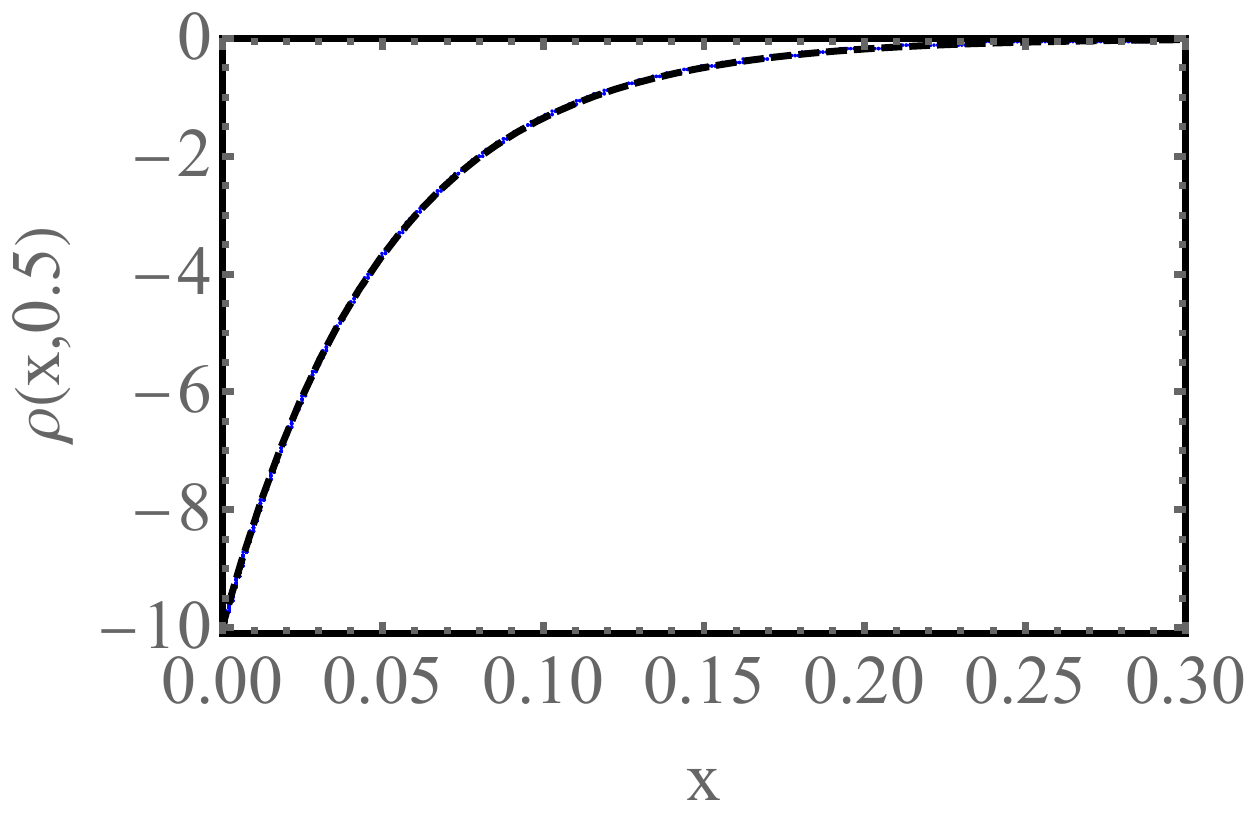}
\includegraphics[width=0.3\textwidth,clip=]{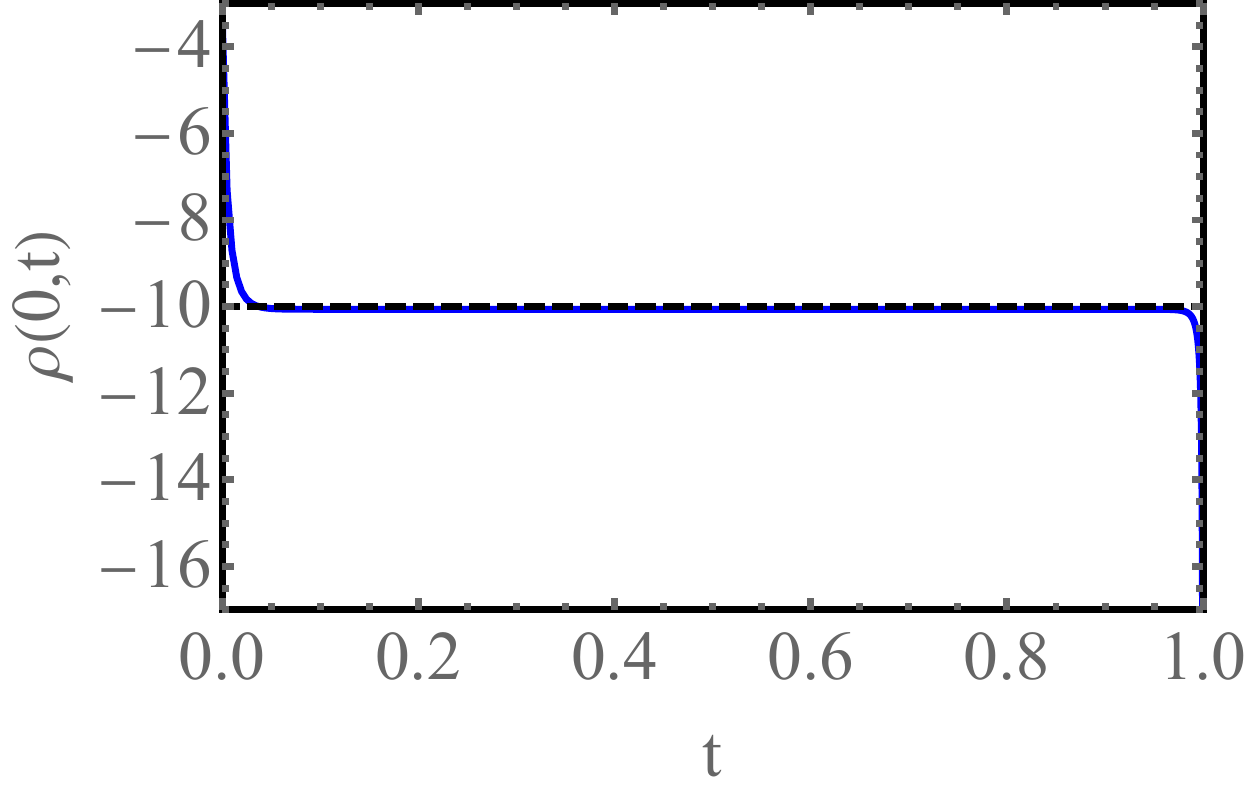}
\caption{Upper panel: the large-$a$ steady-state solution $\rho_{\text{st}}(x)$, predicted by Eq.~(\ref{rho1steady}) (dashed line), is compared to numerically computed $\rho(x,t=0.5)$ (solid line)  for $a=20$ and $\Lambda=-0.5$. The two curves are indistinguishable. Lower panel: numerically computed $\rho(x=0,t)$ versus time (solid line) and theoretical value $\rho(x=0)= \Lambda a=-10$ (dashed line) for the same parameters. The numerical solution exhibits two narrow boundary layers: at $t=0$ and $t=1$, where the steady-state solution~(\ref{rho1steady}) does not apply.}
\label{20linprof}
\end{figure}

Similarly to the full-line problem \cite{MKV,KMSparabola}, the cumulants of the height distribution ${\mathcal P}(H,T,A)$ can be calculated via a regular perturbation theory applied to the OFM problem. The small parameter is $H-H_0(a)$, or $\Lambda$. We set
\begin{eqnarray}
 \!\!\! h(x,t)&=& h_0(x,t)+\Lambda h_1(x,t)+\Lambda^2 h_2(x,t) +\dots ,\label{hexpansion}\\
 \!\!\! \rho(x,t) &=& \Lambda \rho_1(x,t)+\Lambda^2 \rho_2(x,t) +\dots . \label{pexpansion}
\end{eqnarray}
where $h_0(x,t)$ is given by Eq.~(\ref{hdet}). Correspondingly, $s(\Lambda)=\Lambda^2 s_1 +\Lambda^3 s_2 + \dots$. The distribution variance is obtained in the first order of this perturbation series \cite{MKV,KMSparabola}. Here
Eqs.~(\ref{eqh}) and (\ref{eqrho})  yield
\begin{eqnarray}
  && \partial_t h_1 +V_0 \,\partial_x h_1 - \partial_x^2 h_1 = \rho_1, \label{1}\\
  && \partial_t \rho_1 +\partial_x (V_0 \,\rho_1)+\partial_x^2 \rho_1= 0, \label{2}
\end{eqnarray}
and $V_0=V_0(x,t)$ is given by Eq.~(\ref{Vdet}). The boundary conditions are
\begin{eqnarray}
\label{BClin}
&& \partial_x h_1(0,t)=0,\quad \partial_x \rho_1(0,t)+a \rho(0,t)=0, \nonumber \\
&& h_1(x,0)=0, \quad\mbox{and} \quad \rho_1(0,1)=\delta(x).
\end{eqnarray}
The KPZ nonlinearity is at work already in the first  order of the perturbation expansion, so the variance differs from that for the Edwards-Wilkinson equation. Importantly, Eqs.~(\ref{1})-(\ref{BClin}) include only one parameter $a$. The first-order action $s_1$ can therefore depend only on $a$, and the resulting action, corresponding to typical, small fluctuations of height, must scale as
\begin{equation}\label{linearactiongen}
s(H,a) = f(a) \left[H-H_0(a)\right]^2 .
\end{equation}
In order to find the function $f(a)$, one should solve Eqs.~(\ref{1}) and (\ref{2}). However, in spite of their linearity, these equations are hard to solve, because $V_0(x,t)$ depends on $x$ and $t$ in a complicated way, see Eq.~(\ref{Vdet}).  Here we will only consider the limit of $a\gg 1$. In the dimensional variables, this limit corresponds to a \emph{very strong} left inequality in Eq.~(\ref{interm}). In this case $V_0(x,t)\simeq a = \text{const}$ behind the shock, see Fig. \ref{shock}, and Eq.~(\ref{2}) in this region becomes very simple:
\begin{equation}\label{rho1simple}
 \partial_t \rho_1 +a \,\partial_x \rho_1+\partial_x^2 \rho_1= 0.
\end{equation}
As $a\gg 1$, the solution rapidly approaches a steady state, $\rho^{\text{st}}_1 (x)$. By virtue of the boundary condition (\ref{BCrho}), this steady state must have zero flux, and we obtain
\begin{equation}\label{rho1steady}
\rho^{\text{st}}_1 (x) = \Lambda a e^{-ax}\,,
\end{equation}
with the coefficient determined by the conservation law $\int_0^{\infty} \rho_1(x)\,dx = \Lambda$. The solution is strongly localized at $x=0$. Using Eq.~(\ref{eq:sdyn_recast}), we obtain $s\simeq a\Lambda^2/4$. Using the relation \cite{Smithshort}
\begin{equation}\label{Smithrelation}
\frac{ds}{d\Lambda} = \Lambda\,\frac{dH}{d\Lambda} ,
\end{equation}
we express $\Lambda$ via $H$: $\Lambda=(2/a) (H-H_0)$, where the expected height $H_0=H_0(a)$ is given by the first line of Eq.~(\ref{H0asymp}). Finally,
\begin{equation}\label{varlargeA}
s(H,a)= \frac{1}{a}\,\left(H+\frac{a^2}{2}+2 \ln 2 +\dots\right)^2 ,
\end{equation}
so $f(a)=1/a$ in this limit. As to be expected, small fluctuations of the height are normally distributed. The variance of $\mathcal{P}(H)$ is proportional to $a$ in this regime of $a\gg 1$.  In the dimensional variables, the variance is 
\begin{equation}\label{variance}
\text{Var}_H = \frac{A|\lambda|^3 DT}{2\nu^3}.
\end{equation}
That is, for a sufficiently large positive $A$, the customary $\lambda$-independent Edwards-Wilkinson scaling $\text{Var}_H \sim T^{1/2}$, observed at very short times, gives way to a different scaling,  $\text{Var}_H\sim T$, at intermediate times.
Figure \ref{20linear}  compares Eq.~(\ref{varlargeA}) with results of numerical solution of the full OFM problem for $a=20$ and relatively small $|H-H_0(a)|$. A very good agreement is observed.
The steady-state solution (\ref{rho1steady}) does not apply very close to $t=0$ and $t=1$ (see Fig. \ref{20linprof}), but these ``boundary layers in time" would give only a subleading correction (with respect to the small parameter $1/a$) to the action.

\subsection{$\lambda H\to \infty$ tail}
\label{negativetail}
For $a\to 0$ the optimal realization of noise  $\rho$, which determines the $H\to -\infty$ tail of $\mathcal{P}(H)$ (to remind the reader, $\lambda<0$), is strongly localized at $x=0$ and has the form of
(the right half of) a standing soliton which, for the full-line problem, has been known for a long time \cite{Mikhailov1991,Fogedby1998,KK2007,MKV}:
\begin{equation}\label{soliton0rho}
\rho(x,a=0) = -2 |H|\,
   \text{sech}^2\left(\sqrt{\frac{|H|}{2}} x\right) ,\quad x\geq 0.
\end{equation}
The action (\ref{eq:sdyn_recast}), evaluated with this $\rho(x)$,
gives the leading term in the last line of Eq.~(\ref{forreference}). The corresponding optimal
interface slope is
\begin{equation}\label{soliton0V}
V(x,a=0) = \sqrt{2 |H|}\,
   \text{tanh}\left(\sqrt{\frac{|H|}{2}} x\right)
\end{equation}
at $0\leq x< t\,\sqrt{|H|/2}$, and $V=0$ at $x> t\,\sqrt{|H|/2}$.

\begin{figure}[ht]
\includegraphics[width=0.3\textwidth,clip=]{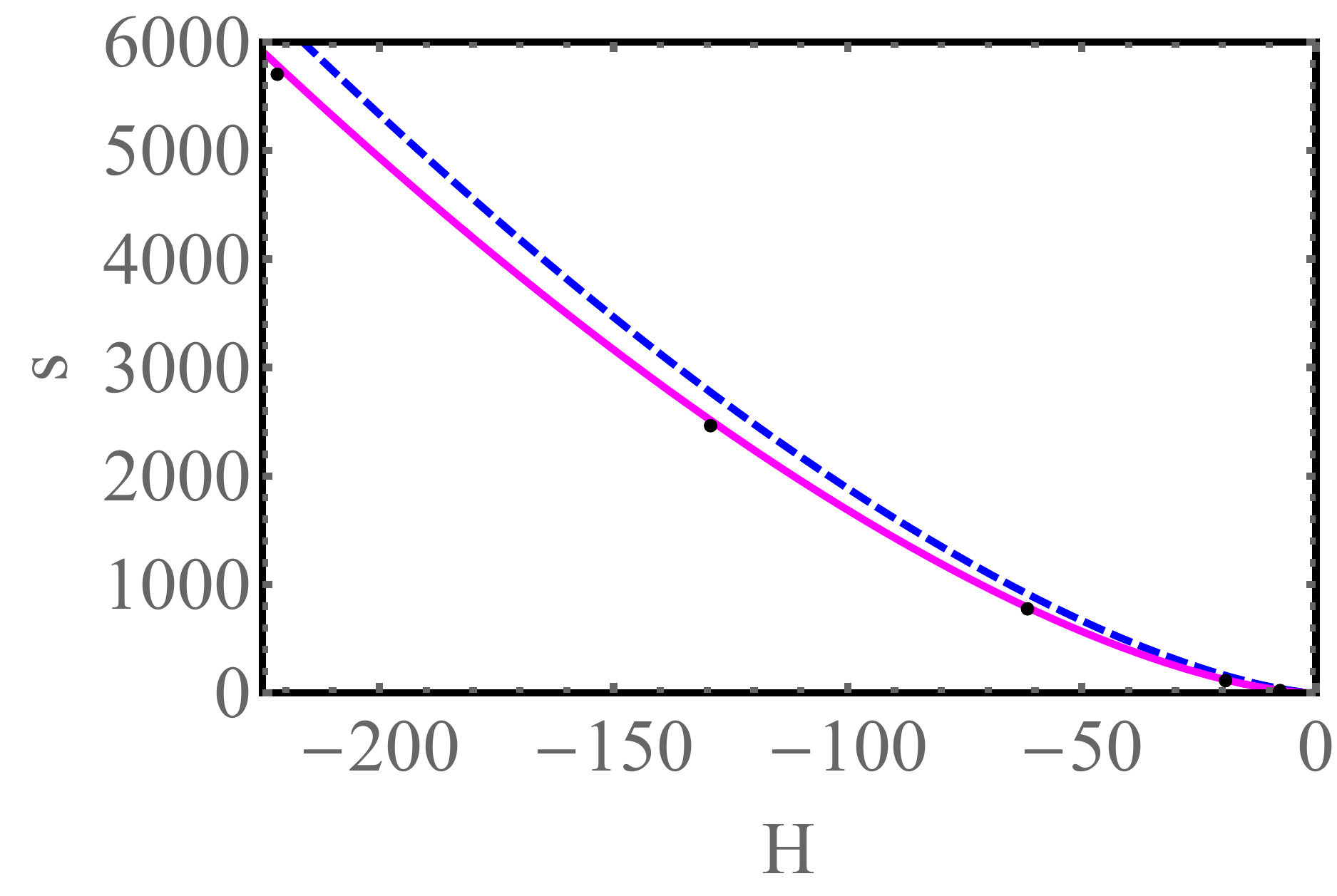}
\includegraphics[width=0.3\textwidth,clip=]{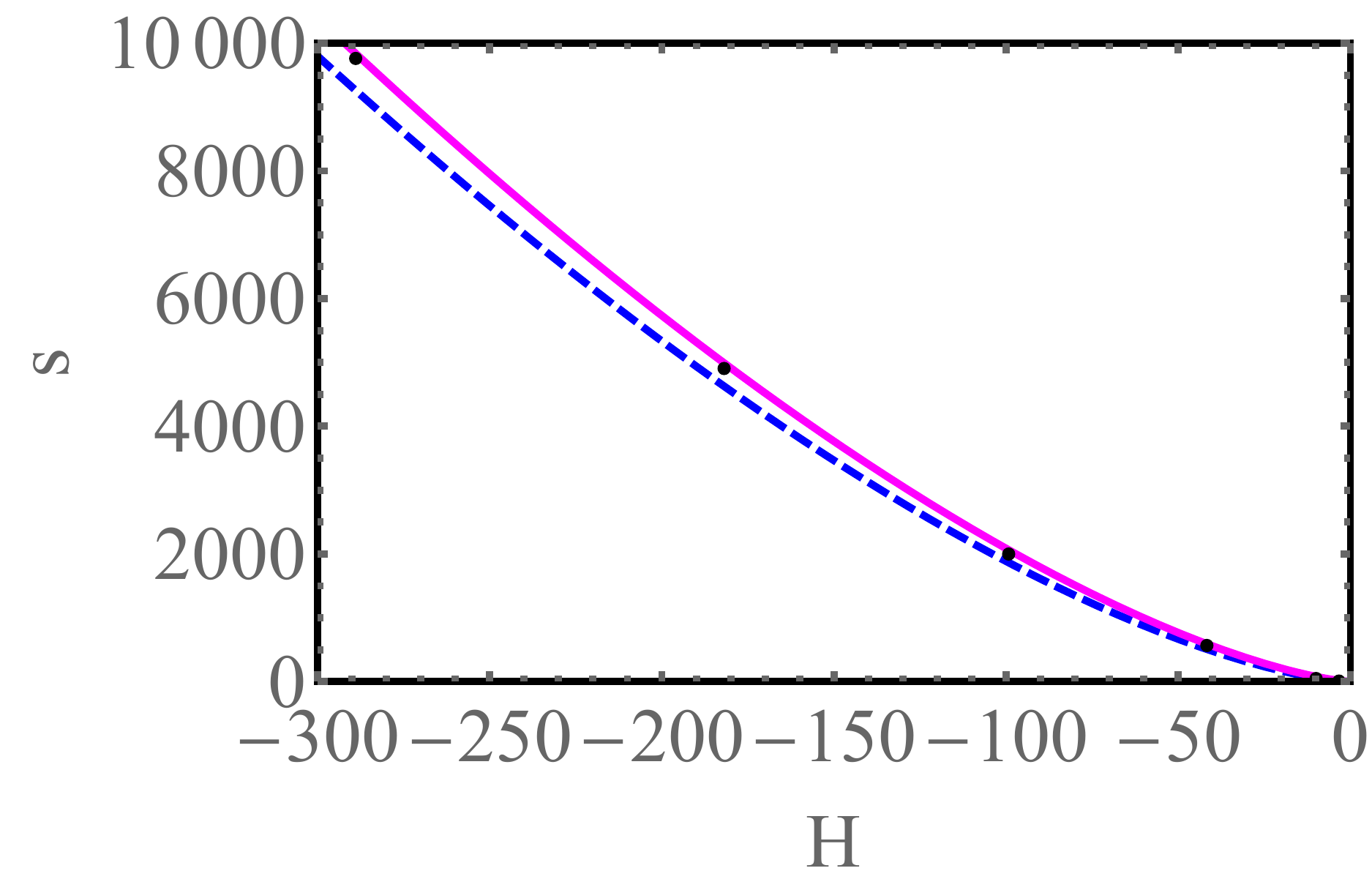}
\caption{The action $s(H,a)$ for large negative $H$ and $a=1$ (upper panel) and $-1$ (lower panel). Solid line: prediction of Eq.~(\ref{sshort}), including the subleading term $-2a|H|$.  Dashed line:
only the leading term in Eq.~(\ref{sshort}). Symbols:  numerical results.}
\label{actsolitonfig}
\end{figure}

\begin{figure}[ht]
\includegraphics[width=0.3\textwidth,clip=]{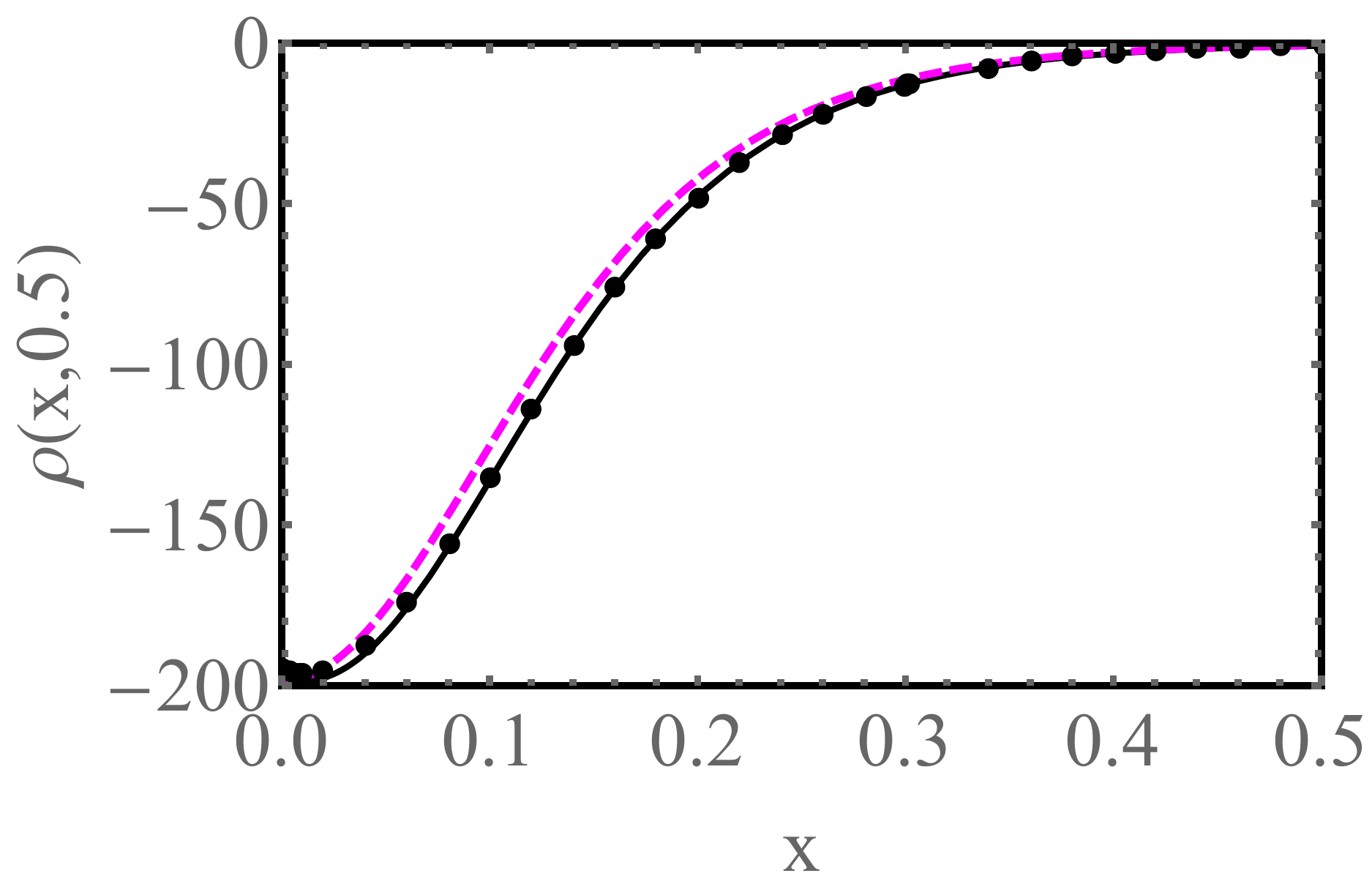}
\caption{The optimal noise realization $\rho(x,t=1/2)$ versus $x$ for $H\simeq -99.3$ and $a=-1$. Solid line: Eq.~(\ref{soliton0rho}) with $\beta=\sqrt{2|H|}$ and $\ell = (2/\beta) \,
\text{arctanh}\,(a/\beta) \simeq A/|H|$. Dashed line: unshifted soliton (\ref{soliton0rho}). Symbols: numerical results.}
\label{shiftsolfig}
\end{figure}

Expression~(\ref{soliton0V}) vanishes at $x=0$ as it should.  It is natural
to assume that the optimal solution for $a\neq 0$ is given
by the $x\geq 0$ part of a standing soliton, shifted along the $x$-axis:
\begin{eqnarray}
 \rho(x,a) &=& -\beta ^2 \text{sech}^2\left[\frac{\beta }{2}
    (x+\ell)\right] , \label{shiftedrho} \\
  V(x,a) &=& \beta  \tanh \left[\frac{\beta}{2}
   (x+\ell)\right] , \label{shiftedV}
\end{eqnarray}
where $\beta$ and $\ell$ should be determined by $H$ and $a$. The boundary condition (\ref{AA})
yields
\begin{equation}\label{ell}
\ell=\frac{2}{\beta}\, \text{arctanh}\left(\frac{a}{\beta}\right),
\end{equation}
and the no-flux boundary condition (\ref{BCrho}) is satisfied automatically. Now we use Eq.~(\ref{eqh}) at $x=0$ to evaluate $\partial_t h(0,t)$:
$$
\partial_t h(0,t) = \partial_x V(0,t)-\frac{a^2}{2}+\rho(0,t).
$$
For the solution~(\ref{shiftedrho}) and (\ref{shiftedV}) the right hand side evaluates to $-\beta^2/2$. This must be equal to $H$, so $\beta=\sqrt{2|H|}$, as in the case of $a=0$.
Now the shifted solution solution is fully determined. Evaluating its action~(\ref{eq:sdyn_recast}),
we obtain
\begin{equation}\label{slong}
s(H,a)=\frac{4\sqrt{2}}{3} \left|H\right|^{3/2}-2 a |H|+ \frac{a^3}{3} .
\end{equation}
The first term is the leading one. It coincides with its counterpart for $a=0$, see the last line of Eq.~(\ref{forreference}).  The last term in Eq.~(\ref{slong}) appears to be in excess of accuracy. This is because it is much smaller, at large $|H|$,  than subleading terms unaccounted for by the soliton solution. Indeed, for $A=0$ the subleading term in the asymptotic expansion of the exact large-deviation function $s$ at large negative $H$ scales as $|H|^{1/2}$, see the last line of Eq.~(\ref{forreference}), so it is much larger than $a^3/3$. The second term in Eq.~(\ref{sshort}) scales as $|H|$, and we argue that it is a correct subleading term. We finally obtain
\begin{equation}\label{sshort}
s(H,a)\simeq \frac{4\sqrt{2}}{3} \left|H\right|^{3/2}-2 a |H|,\quad -H\to \infty.
\end{equation}
The asymptotic~(\ref{sshort}), including the subleading term, agrees with our numerics, see  Fig. \ref{actsolitonfig}.
We also verified numerically that the optimal noise realization $\rho(x,t)$ does not change in time except very close to $t=0$ and $t=1$. In addition, it is described well by the shifted soliton solution~(\ref{shiftedrho}), see Fig. \ref{shiftsolfig}.

We emphasize that, in view of Eq.~(\ref{AA}),  the subleading term $-2 a |H|$ of Eq.~(\ref{sshort}) is time-dependent in the dimensional variables, and so it violates the simple scaling $-\ln \mathcal{P}\left(H,t\right)\simeq s \left(H\right)/\sqrt{t}$.

\subsection{$\lambda H \to -\infty$ tail}
\label{positivetail}

On an infinite line the optimal history of $h$ and $\rho$ at $H \to \infty$ is approximately described by a combination of two hydrodynamic solutions, obtained when neglecting the diffusion terms in Eqs.~(\ref{eqh}) and (\ref{eqrho}). The first of the solutions solves the equations \cite{KK2007,KK2009,MKV}
\begin{eqnarray}
 \partial_t \rho +\partial_x (\rho V)&=& 0, \label{rhoeq}\\
  \partial_t V +V \partial_x V &=&\partial_x \rho, \label{Veq}
\end{eqnarray}
with $\rho(x,t)\neq 0$. These equations describe a non-stationary inviscid flow of an effective gas with density $\rho$ and velocity $V$. The gas pressure is \emph{negative}: $p(\rho)=-\rho^2/2$.
With the initial condition $V(x,t)=0$ and the final condition (\ref{pT}), the solution represents a uniform-strain flow on a shrinking finite support \cite{KK2009,MKV} which leads to collapse of the gas into the origin at $t=1$.

The second solution appears in the regions where $\rho(x,t)=0$. It solves the Hopf equation $\partial_tV+V\partial_xV=0$. The two hydrodynamic solutions can be continuously matched \cite{MKV}. The $H\to \infty$ tail of $\mathcal{P}(H)$ is determined solely by the ``pressure-driven" solution, and one arrives at \cite{KK2009,MKV}
\begin{equation}\label{HDinfinite}
s= \frac{8\sqrt{2}}{15\pi}\,H^{5/2} .
\end{equation}
For the half-line problem with $A=0$ the optimal path is given by the right half of the full-line solution. The resulting action is twice as small as for the full line, and we arrive at the leading term of the first line in Eq.~(\ref{forreference}).

\begin{figure}[h]
\includegraphics[width=0.3\textwidth,clip=]{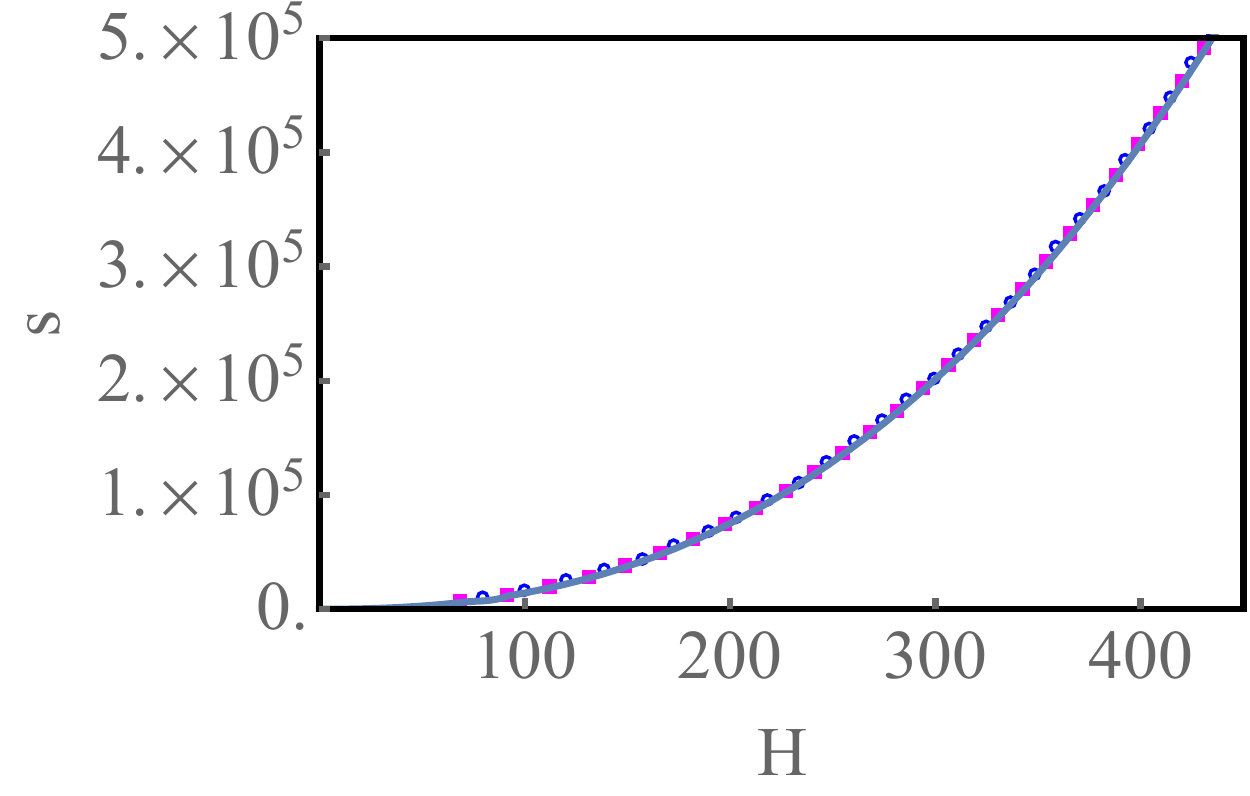}
\caption{The action $s(H)$ for large positive $H$ and $a=0$ (solid line), $a=1$ (circles) and $a=-1$ (rectangles). The solid line actually represents two indistinguishable lines which show the analytic expression from Ref. \cite{SM2018} and the numerical results from Ref. \cite{MKV}.}
\label{actionHD}
\end{figure}

When $a\neq 0$, the ``gas velocity" at $x=0$ is fixed and nonzero: $V(0,t)=a$.   To get an insight into the character of solution, let us perform the ``hydrodynamic" rescaling \cite{MKV} of the original OFM equations and boundary conditions: $x/\Lambda^{1/3}\to x$, $V/\Lambda^{1/3}\to V$, and $\rho\to \rho/\Lambda^{2/3}$. In the new variables the OFM equations become
\begin{eqnarray}
 \partial_t \rho +\partial_x (\rho V)&=& \frac{1}{\Lambda^{2/3}} \,\partial_x^2 V , \label{rhoeqHD}\\
  \partial_t V +V \partial_x V -\partial_x \rho &=&-\frac{1}{\Lambda^{2/3}}\, \partial_x^2 \rho .\label{VeqHD}
\end{eqnarray}
The boundary conditions in time are $V(x,t)=0$ and $\rho(x,1)=\delta(x)$, whereas the boundary conditions at $x=0$
become
\begin{eqnarray}
  &&V(0,t) = \frac{a}{\Lambda^{1/3}} , \label{BCLambdaV}\\
  && a \rho(0,t) = - \frac{1}{\Lambda^{1/3}}\,\partial_x\rho(0,t). \label{BCLambdarho}
\end{eqnarray}
When $\Lambda \to \infty$, which corresponds to $H\to \infty$,  we can drop the diffusion terms in Eqs.~(\ref{rhoeqHD}) and (\ref{VeqHD}) thus reproducing Eqs.~(\ref{rhoeq}) and (\ref{Veq}). This procedure applies in the large hydrodynamic region outside of two narrow boundary layers. The first boundary layer appears,  for any $a$, between the ``pressure-driven" flow region and the Hopf flow region. The second boundary layer, at $x=0$, appears only when $a\neq 0$. These two boundary layers give only subleading contributions to the action. To calculate the leading term it suffices to use the hydrodynamic equations (\ref{rhoeq}) and (\ref{Veq}).  As their order is reduced compared with the full equations, one is allowed to use only one of the two boundary conditions~(\ref{BCLambdaV}) and (\ref{BCLambdarho}). It is convenient to use Eq.~(\ref{BCLambdaV}) which, as $\Lambda$ goes to infinity,  has a simple limit $V(x,t)=0$. The ensuing hydrodynamic problem is independent of $a$, and its solution is described by the right half of the full-line hydrodynamic solution \cite{KK2009,MKV}. This leads us to the conclusion that the leading-order action, for any finite $a$, is described by the first term in the first line of Eq.~(\ref{forreference}).

\begin{figure} [ht]
\includegraphics[width=0.3\textwidth,clip=]{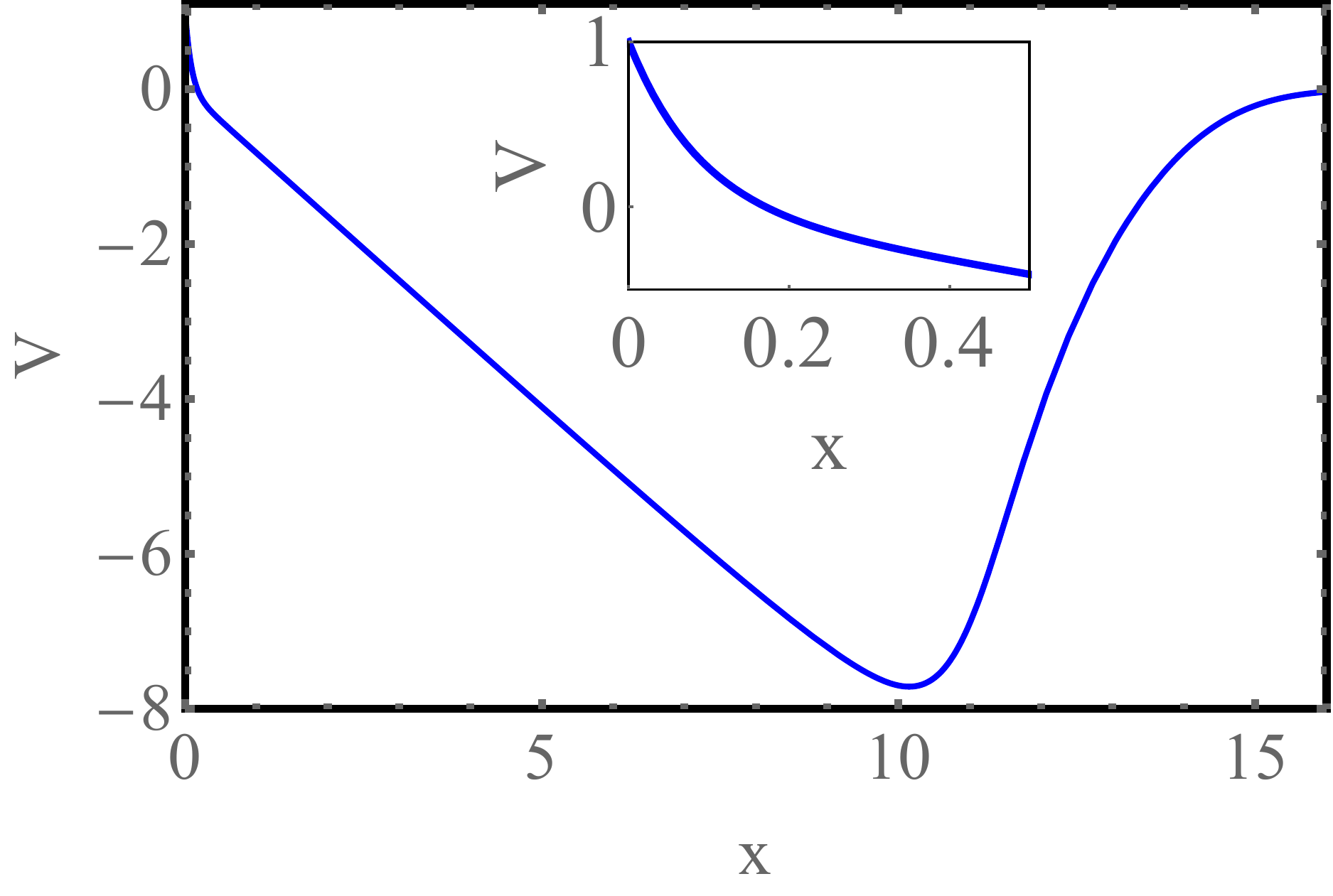}
\includegraphics[width=0.3\textwidth,clip=]{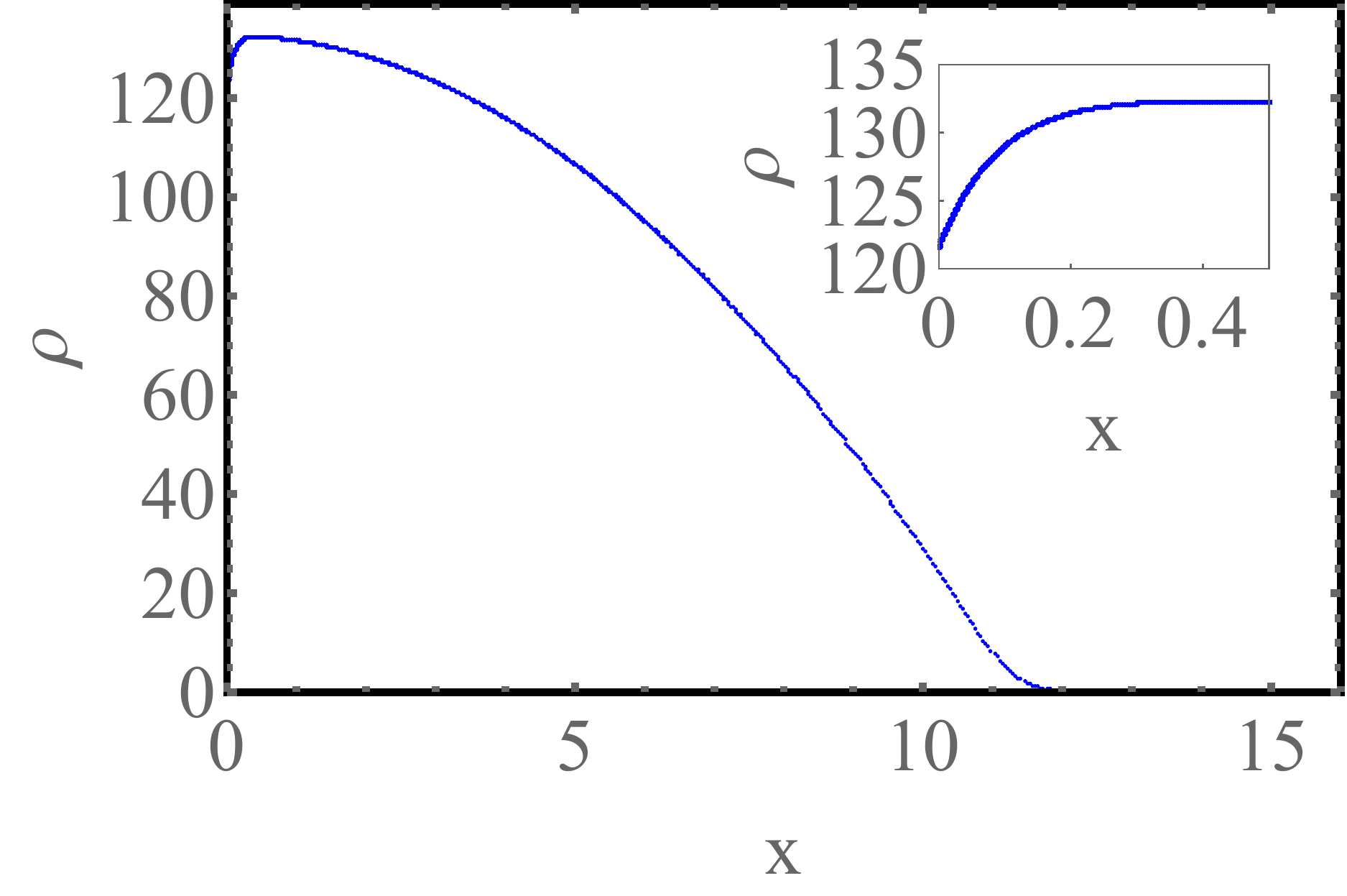}
\includegraphics[width=0.3\textwidth,clip=]{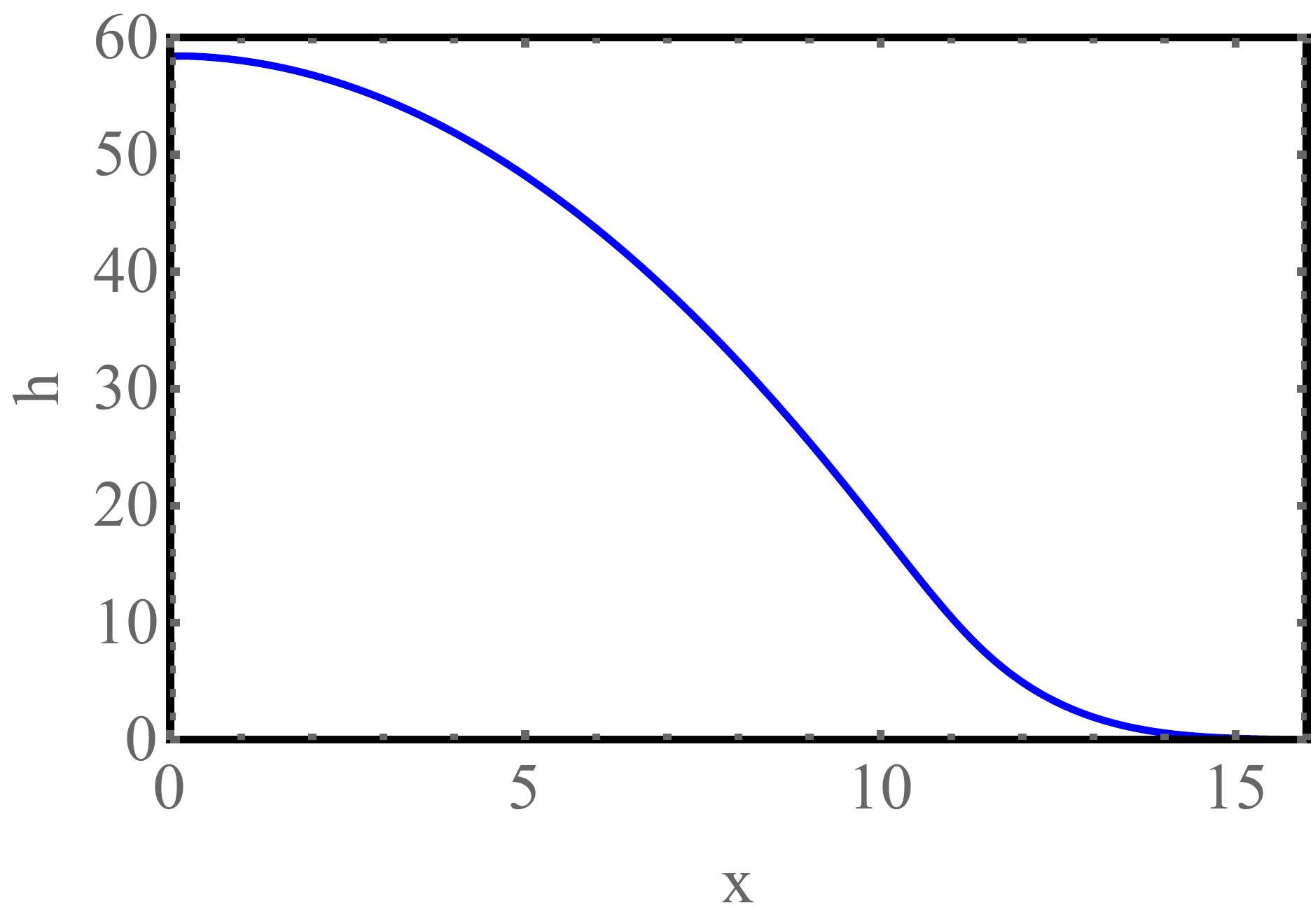}
\caption{The $x$-profiles of $V$, $\rho$ and $h$  at $t=1/2$ for $a=1$ and $\Lambda=10^3$. The corresponding  value of $H=212.9$. The insets show blowups of the boundary layer at $x=0$.}
\label{HDprofiles1}
\end{figure}

\begin{figure} [ht]
\includegraphics[width=0.3\textwidth,clip=]{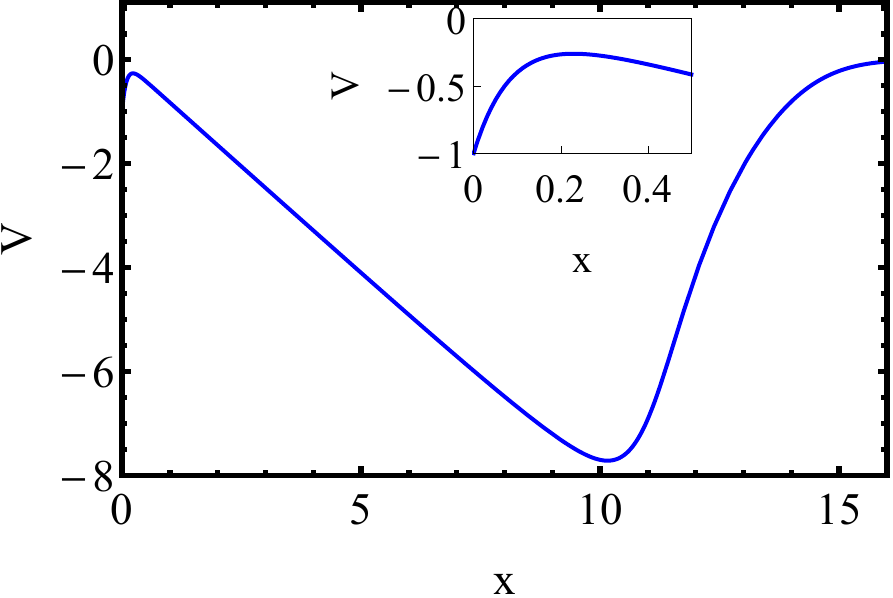}
\includegraphics[width=0.3\textwidth,clip=]{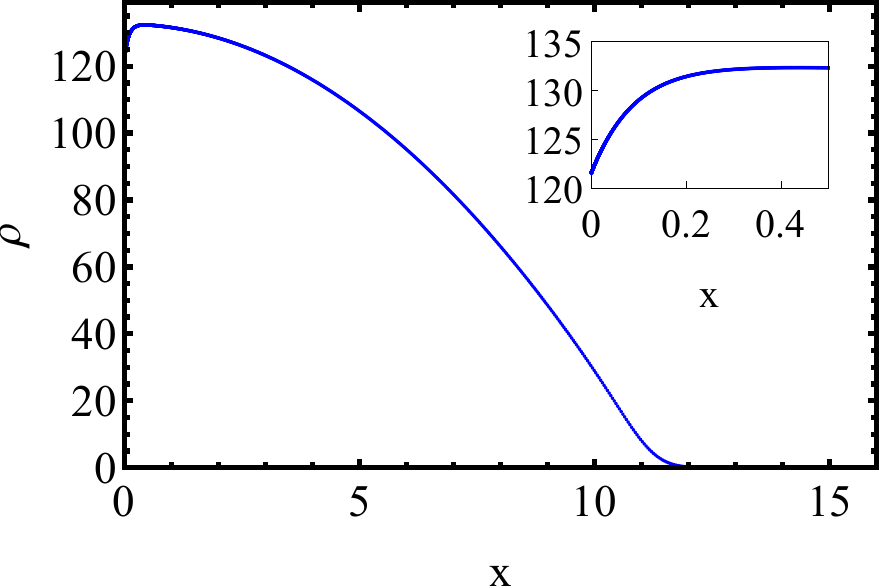}
\includegraphics[width=0.3\textwidth,clip=]{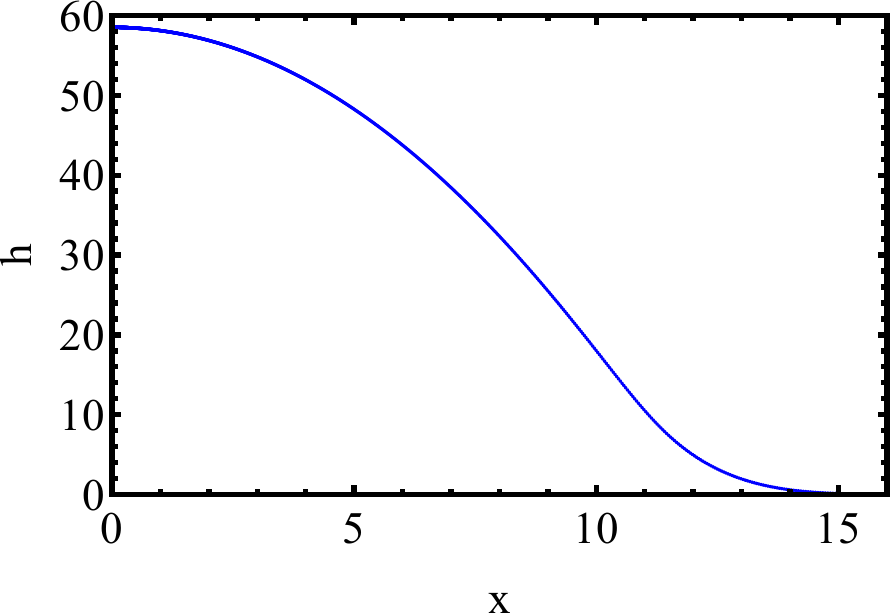}
\caption{The $x$-profiles of $V$, $\rho$ and $h$ at $t=1/2$ for $a=-1$ and $\Lambda=10^3$. The corresponding  value of $H=213.2$. The insets show blowups of the boundary layer at $x=0$.}
\label{HDprofilesm1}
\end{figure}

We verified this important result numerically. Figure \ref{actionHD} shows that $s(H)$ for large positive $H$ almost coincide for $a=0$, $1$ and $-1$. The profiles of $V(x,t=0.5)$, $\rho(x,t=0.5)$ and $h(x,t=0.5)$ are shown
in Figs. \ref{HDprofiles1} and \ref{HDprofilesm1} for $a=1$ and $-1$, respectively. One can see that, outside of a narrow
boundary layer at $x=0$ the solution is large-scale and close to the analytical solution for $a=0$, obtained in Refs. \cite{KK2009,MKV}. The presently unknown subleading
terms, coming from the boundary layer at $x=0$, will violate the simple scaling $-\ln \mathcal{P}\simeq s(H)/\sqrt{t}$.

\section{Summary and discussion}
\label{disc}

We studied, analytically and numerically,  the short-time distribution of the height of the KPZ interface on a half-line with a prescribed interface slope $A$ at $x=0$ for flat initial condition. We showed that, for small and moderate slopes, one observes a simple $A$-independent scaling $-\ln \mathcal{P}\left(H,A,t\right)\simeq s(H)/\sqrt{t}$, and  $s(H)$
obeys the simple relation~(\ref{relation}).

For sufficiently large slopes, there are two regimes: of very short times and of intermediate times. At very short times the simple $A$-independent scaling $-\ln \mathcal{P}\left(H,A,t\right)\simeq s(H)/\sqrt{t}$ is observed, and a finite $A$ is irrelevant. For intermediate times the scaling behavior of  $-\ln \mathcal{P}\left(H,A,t\right)$ is more general, see Eq.~(\ref{newscaling}). The more general scaling is most
pronounced in the body of the height distribution, see \textit{e.g.} Eq.~(\ref{variance}).   The positive and negative tails do not depend on $A\sqrt{t}$ in the leading order, and they obey Eq. ~(\ref{relation}). In a subleading order the simple scaling is violated. We calculated one such scaling-violating subleading term in the $\lambda H \to \infty$ tail,
see Eq.~(\ref{sshort}).

The $\lambda H \to \infty$ tail is quite universal. Indeed, slightly extending a previous argument for infinite systems \cite{KMSparabola}, one realizes that Eq.~(\ref{sshort}), including its subleading term, is valid for a whole class of initial conditions. The reason is that, as $\lambda H\to \infty$, the slightly shifted soliton solution~(\ref{shiftedrho}) and (\ref{shiftedV}) is strongly localized near the boundary $x=0$ and therefore is not sensitive to specifics of the (deterministic) initial condition. At the level of the leading-order description of this tail, an immediate confirmation comes from the droplet case \cite{Krajenbrink2018}, for two different finite values of $A$.

The $\lambda H\to -\infty$ tail is not as universal. Here the numerical factor, multiplying the leading term $H^{5/2}$ of $s(H)$,  should depends on the initial condition,  as already observed in infinite systems \cite{KK2009,MKV,KMSparabola,DMRS}.

Finally, based on the previous work for infinite systems \cite{SMP,Corwinetal2018}, one can expect that both tails of the short-time height distribution, $4\sqrt{2}|H|^{3/2}/3$ and $4\sqrt{2}H^{5/2}/(15\pi)$, continue to hold, at sufficiently large $|H|$, at all times.

\section*{ACKNOWLEDGMENTS}

We are grateful to Naftali Smith for a valuable advice and a critical reading of the manuscript. B.M. acknowledges financial support from the Israel Science Foundation (grant No. 807/16).


\bigskip\bigskip

\begin{thebibliography}{99}


\bibitem{KPZ}  M. Kardar, G. Parisi, and Y.-C. Zhang, Phys. Rev. Lett. \textbf{56}, 889 (1986).

\bibitem{signlambda}
Changing the sign of $\lambda$ is equivalent to changing $h$ to $-h$.


\bibitem{HHZ} T. Halpin-Healy and Y.-C. Zhang, Phys. Reports \textbf{254}, 215 (1995); T. Halpin-Healy and K. A. Takeuchi,
J. Stat. Phys. \textbf{160}, 794 (2015).

\bibitem{Barabasi} A.-L. Barabasi and H. E. Stanley, {\it Fractal Concepts in Surface Growth} (Cambridge
University Press, Cambridge, UK, 1995).

\bibitem{Krug}
J. Krug, Adv. Phys. \textbf{46}, 139 (1997).

\bibitem{Corwin}
I. Corwin, Random Matrices: Theory Appl. \textbf{1}, 1130001 (2012).

\bibitem{QS}
J. Quastel and  H. Spohn, J. Stat. Phys. \textbf{160}, 965 (2015).

\bibitem{S2016}
H. Spohn, in \textit{Stochastic Processes and Random Matrices}, \textit{Lecture Notes of the Les Houches Summer School},  edited by G. Schehr, A. Altland, Y. V. Fyodorov and L. F. Cugliandolo (Oxford University Press, Oxford, 2015), vol. 104.

\bibitem{Takeuchi2017} K. A. Takeuchi, Physica A \textbf{504}, 77 (2018).

\bibitem{DMRS} P. Le Doussal, S. N. Majumdar, A. Rosso, and G. Schehr,
Phys. Rev. Lett. \textbf{117}, 070403 (2016).

\bibitem{LeDoussal2017} A. Krajenbrink and P. Le Doussal, Phys. Rev. E \textbf{96}, 020102(R) (2017).

\bibitem{SM2018}  N. R. Smith and B. Meerson, Phys. Rev. E \textbf{97}, 052110 (2018).

\bibitem{MKV} B. Meerson, E. Katzav, and A. Vilenkin, Phys. Rev. Lett. \textbf{116}, 070601 (2016).


\bibitem{SMP} P. V. Sasorov, B. Meerson, and S. Prolhac,
J. Stat. Mech. (2017) P063203.

\bibitem{Corwinetal2018} I. Corwin, P. Ghosal, A. Krajenbrink,  P. Le Doussal, and L.-C. Tsai, 
arXiv:1803.05887.

\bibitem{footnote:displacement}
We subtract from $h(0,t)$ the \emph{noise-induced} systematic ``drift" of the KPZ interface.

\bibitem{SMS2018} N. R. Smith, B. Meerson and P. V. Sasorov, J. Stat. Mech. (2018) 023202.


\bibitem{GueudreLeDoussal2012} T. Gueudr\'{e} and P. Le Doussal, Europhys. Lett. \textbf{100}, 26006  (2012).

\bibitem{Borodin2016} A. Borodin, A. Bufetov, and I. Corwin, Annals of Phys. \textbf{368}, 191 (2016).

\bibitem{Barraquand2017} G. Barraquand, A. Borodin, I. Corwin, and M. Wheeler, arXiv:1704.04309.

\bibitem{CorwinShen2018} I. Corwin and H. Shen, 
arXiv:1610.04931.

\bibitem{ItoTakeuchi2018} Y. Ito and K. A. Takeuchi, Phys. Rev. E \textbf{97}, 040103(R) (2018).



\bibitem{Krajenbrink2018} A. Krajenbrink and P. Le Doussal, arXiv:1804.08800.

\bibitem{naive} One would na\"{\i}vely think that, as an infinite line provides more degrees of freedom than a half-line, the system will utilize them when conditioned on an unusual value of the height $H$.
It may come as a surprise that the system often prefers \emph{not} to explore so many degrees of freedom (and, in particular, not to break the mirror symmetry $x \leftrightarrow -x$ of the optimal path).

\bibitem{Kardar} M. Kardar, Phys. Rev. Lett. \textbf{55}, 2235 (1985).


\bibitem{Halperin} B. I. Halperin and M. Lax, Phys. Rev. \textbf{148}, 722 (1966).

\bibitem{Langer} J. Zittartz and J. S. Langer, Phys. Rev. \textbf{148}, 741 (1966).

\bibitem{Lifshitz} I. M. Lifshitz, Zh. Eksp. Teor. Fiz. \textbf{53}, 743 (1967) [Sov. Phys.
JETP \textbf{26}, 462 (1968)].
\bibitem{Lifshitz1988} I. Lifshitz, S. Gredeskul, and A. Pastur, {\it Introduction to the Theory of Disordered Systems} (Wiley, New York, 1988).

\bibitem{turb1} G. Falkovich, I. Kolokolov, V. Lebedev, and A. Migdal, Phys. Rev. E \textbf{54}, 4896 (1996).

\bibitem{turb2} G. Falkovich, K. Gaw\c{e}dzki, and M. Vergassola, Rev. Mod. Phys. \textbf{73}, 913 (2001).

\bibitem{turb3} T. Grafke, R. Grauer, and T. Sch\"{a}fer, J. Phys. A \textbf{48},  333001 (2015).

\bibitem{bertini2015} L. Bertini, A. De Sole, D. Gabrielli, G. Jona-Lasinio, and C. Landim, Rev. Mod. Phys. \textbf{87}, 593 (2015).

\bibitem{EK}  V. Elgart and A. Kamenev, Phys. Rev. E \textbf{70}, 041106 (2004).

\bibitem{MS2011} B. Meerson and P.V. Sasorov, Phys. Rev. E \textbf{83}, 011129 (2011); \textbf{84}, 030101(R) (2011).

\bibitem{Mikhailov1991} A. S. Mikhailov, J. Phys. A \textbf{24}, L757 (1991).

\bibitem{GurarieMigdal1996} V. Gurarie and A. Migdal, Phys. Rev. E \textbf{54}, 4908 (1996).

\bibitem{Fogedby1998} H.C. Fogedby, Phys. Rev. E \textbf{57}, 4943 (1998).

\bibitem{Fogedby1999} H.C. Fogedby, Phys. Rev. E \textbf{59}, 5065 (1999).

\bibitem{Nakao2003} H. Nakao and A. S. Mikhailov, Chaos \textbf{13}, 953 (2003).


\bibitem{KK2007} I. V. Kolokolov and S. E. Korshunov, Phys. Rev. B \textbf{75}, 140201(R) (2007).

\bibitem{KK2008} I. V. Kolokolov and S. E. Korshunov, Phys. Rev. B \textbf{78}, 024206 (2008).

\bibitem{KK2009} I. V. Kolokolov and S. E. Korshunov, Phys. Rev. B \textbf{80}, 031107 (2009).

\bibitem{Fogedby2009} H.C. Fogedby and W. Ren, Phys. Rev. E \textbf{80}, 041116 (2009).

\bibitem{KMSparabola} A. Kamenev, B. Meerson, and P. V. Sasorov, Phys. Rev. E \textbf{94}, 032108 (2016).

\bibitem{Janas2016} M. Janas, A. Kamenev, and B. Meerson, Phys. Rev. E \textbf{94}, 032133 (2016).

\bibitem{MeersonSchmidt2017} B. Meerson and J. Schmidt, J. Stat. Mech. (2017) P103207.

\bibitem{MSV_3d} B. Meerson, P. V. Sasorov and A. Vilenkin, J. Stat. Mech. (2018) 053201.

\bibitem{SKM2018} N. R. Smith, A. Kamenev and B. Meerson, Phys. Rev. E \textbf{97}, 042130 (2018).

\bibitem{Whitham} G. B. Whitham, \textit{Linear and Nonlinear Waves} (Wiley, New York,
1974).

\bibitem{Frey1996} E. Frey, U. C. T{\"a}uber, and T. Hwa, Phys. Rev. E \textbf{53}, 4424 (1996).

\bibitem{Canet2011} L. Canet, H. Chat\'{e}, B. Delamotte, and N. Wschebor, Phys. Rev.
E \textbf{84}, 061128 (2011); \textbf{86}, 019904(E) (2012).

\bibitem{Mathey2017} S. Mathey, E. Agoritsas, T. Kloss, V. Lecomte, and L. Canet, Phys. Rev. E \textbf{95}, 032117 (2017).

\bibitem{IS} T. Imamura and T. Sasamoto, Phys. Rev. Lett. \textbf{108}, 190603 (2012); J.
Stat. Phys. \textbf{150}, 908 (2013).

\bibitem{Borodinetal} A. Borodin, I. Corwin, P.L. Ferrari, and B. Vet\H{o}, Mathematical Physics, Analysis and Geometry \textbf{18}, 20 (2015).

\bibitem{numerics}  We solved the OFM equations  numerically with the Chernykh-Stepanov back-and-forth iteration algorithm \citep{Chernykh}.

\bibitem{Smithshort} N. Smith (unpublished).


\bibitem{Chernykh}  A. I. Chernykh and M. G. Stepanov, Phys. Rev. E \textbf{64}, 026306 (2001).

\end{thebibliography}
\end{document}